\documentclass[12pt]{article}

\usepackage{a4wide}
\usepackage{hyperref}
\usepackage{graphicx}
\usepackage{amsmath}
\usepackage{amssymb}
\usepackage{caption}
\newcommand{\GeV}{\,\text{GeV}}
\newcommand{\TeV}{\,\text{TeV}}

\newcommand{\PU}{\text{PU}}
\newcommand{\ntr}{\text{ntr}}

\newcommand{\NpC}{\text{NpC}}
\newcommand{\cut}{\text{cut}}

\date{}

\title{\sf On the use of charged-track information \\to subtract neutral pileup}
\author{Matteo Cacciari,$^{1,2,3}$ Gavin P. Salam$^{4,}$\footnote{On
    leave from CNRS, UMR 7589, LPTHE, F-75005, Paris, France.}\;  and Gregory
  Soyez$^{5}$\\\footnotesize
\footnotesize$^1$Universit\'e Paris Diderot, Paris, France\\
\footnotesize$^2$Sorbonne Universit\'es, UPMC Univ Paris 06, UMR 7589, LPTHE, F-75005, 
Paris, France\\
\footnotesize$^3$CNRS, UMR 7589, LPTHE, F-75005, Paris, France\\
\footnotesize$^4$CERN, PH-TH, CH-1211 Geneva 23, Switzerland\\
\footnotesize\mbox{$^5$IPhT, CEA Saclay, CNRS UMR 3681, F-91191 Gif-sur-Yvette, France}
}

\begin{document}
\maketitle
\vspace{-10cm}
\begin{flushright}
  CERN-PH-TH/2014-052\\
  April 2014\\
  revised February 2015
  %\\
  %\comment{$Rev: 5005 $}
\end{flushright}
\vspace{8cm}

\begin{abstract}
  The use of charged pileup tracks in a jet to predict the neutral
  pileup component in that same jet could potentially lead to improved
  pileup removal techniques, provided there is a strong local
  correlation between charged and neutral pileup.
  In Monte Carlo simulation we find that the correlation is however
  moderate, a feature that we attribute to characteristics of the
  underlying non-perturbative dynamics.
  Consequently, `neutral-proportional-to-charge' (NpC) pileup
  mitigation approaches do not outperform existing, area-based, pileup
  removal methods.
  This finding contrasts with the arguments made in favour of a new
  method, ``jet cleansing'', in part based on the NpC approach.
  We identify the critical differences between the performances of linear
  cleansing and trimmed NpC as being due to the former's
  rejection of subjets that have no charged tracks from
  the leading vertex, a procedure that we name ``zeroing''.
  Zeroing, an extreme version of the ``charged-track trimming''
  proposed by ATLAS, can be combined with a range of pileup-mitigation
  methods, and appears to have both benefits and drawbacks.
  We show how the latter can be straightforwardly alleviated.
  We also discuss the limited potential for improvement that can be
  obtained by linear combinations of the NpC and area-subtraction
  methods.
\end{abstract}

%======================================================================
\section{Introduction}

Pileup, the superposition of many soft proton--proton collisions over
interesting hard-scattering events, is a significant issue at CERN's
Large Hadron Collider (LHC) and also at possible future hadron
colliders.
It affects many observables, including lepton and photon isolation,
missing-energy determination and especially jet observables.
One main technique currently in use to remove pileup from jet
observables~\cite{ATLAS-PU-Performance,CMS-PU-Performance} is
known as the area--median approach~\cite{areasub,areas}.
It makes an event-wide estimate of the pileup level, $\rho$, and then
subtracts an appropriate 4-momentum from each jet based on its area,
i.e.\ its extent in rapidity and azimuth.

Detector-level information can also help mitigate the effect of
pileup: for example, with methods such as particle flow~\cite{pflow}
reconstruction, it is to some extent possible to eliminate the charged
component of pileup, through the subtraction of contributions from
individual charged pileup hadrons.\footnote{Another experimental input
  that could conceivably help reject pileup in future detectors is
  precise timing information.
  One might also wonder about the potential benefit from calorimetric
  pointing information for photons, given that this is already being
  used to locate the primary vertex in Higgs decays to two
  photons~\cite{Aad:2012tfa}. However it seems likely that the
  degradation of pointing angular resolution~\cite{Colas:2005jn} due
  to the lower energy of pileup photons and the higher detector
  occupancy would render this approach impractical. We thank Isabelle
  Wingerter for helpful explanations on this point.}
However, even with such charged hadron subtraction (CHS), there is
always a substantial remaining (largely) neutral pileup contribution,
which remains to be removed.
Currently, when CHS is used, area--median subtraction is then applied
to remove the remaining neutral pileup.

Another approach is to use the information about charged pileup
hadrons in a specific jet to estimate and subtract the remaining
neutral component, without any reference to a jet area or a global
event energy density $\rho$.
Its key assumption is that the neutral energy flow is proportional to
the charged energy flow and so we dub it
neutral-proportional-to-charged (NpC) subtraction.
An advantage that one might imagine for NpC subtraction is that, by
using \emph{local} information about the charged pileup, it might be
better able to account for variations of the pileup from point-to-point 
within the event than methods that rely on event-wide pileup estimates.
We understand that there has been awareness of this kind of approach
in the ATLAS and CMS collaborations for some time now, and we
ourselves also investigated it some years ago~\cite{ourtalk4cms}.
Our main finding was that at particle level it performed marginally
worse than area subtraction combined with CHS.
From discussions with colleagues in the experimental collaborations,
we had the expectation that there might be further degradation at
detector level.
Accordingly we left our results unpublished.

Recently Ref.~\cite{Krohn:2013lba} (KLSW) made a proposal for an
approach to pileup removal named Jet Cleansing.
One of the key ideas that it uses is precisely the NpC
method,\footnote{They also investigated the use of a variable known as
  the jet vertex fraction, widely used experimentally to reject jets
  from a vertex other than the leading
  one~\cite{ATLAS-PU-Performance,CMS:2013wea,ATL-Pileup-2}.}
applied to subjets, much in the way that area--median
subtraction has in the past~\cite{quality,boost2012} been used with
filtering~\cite{filter} and trimming~\cite{trim}.
KLSW found that cleansing brought large improvements over area--median
subtraction. 

Given our earlier findings, KLSW's result surprised us.
The purpose of this article is therefore to revisit our study of the
NpC method and also carry out independent tests of cleansing, both to
examine whether we reproduce the large improvements that they observed
and to identify possible sources of differences.
As part of our study, we will investigate what properties of events
can provide insight into the performance of the NpC method.
We will also be led to discuss the possible value of charged tracks
from the leading vertex in deciding whether to keep or reject
individual subjets (as in charged-track based trimming of
Ref.~\cite{ATL-Pileup-2}).
Finally we shall also examine how one might optimally combine NpC and
area--median subtraction.

%======================================================================
\section{The Neutral-proportional-to-Charged method}
\label{sec:NpC}

NpC subtraction relies on the experiments' ability to identify whether
a given charged track is from a pileup vertex, in order to measure
the charged pileup entering a particular jet.
To a good extent this charged component can be removed, for
example as in CMS's Charged-Hadron Subtraction (CHS) procedure in the
context of particle flow~\cite{pflow}.
The NpC method then further estimates and subtracts the neutral pileup
component by assuming it to be proportional to the charged pileup
component.
At least two variants can be conceived of. 

If the charged pileup particles are kept as part of the jet during
clustering, then the corrected jet momentum is~\cite{ourtalk4cms}
\begin{equation}
  \label{eq:npc-full}
  p_\mu^\text{jet,sub} = p_\mu^\text{jet} -
  \frac{1}{\gamma_0} p_\mu^\text{jet,chg-PU} \,,
\end{equation}
where $p_\mu^\text{jet,chg-PU}$ is the four-momentum of the
charged-pileup particles in the jet and $\gamma_0$ is the
average fraction of pileup transverse momentum that is carried by
charged particles. 
Specifically, one can define
\begin{equation}
  \label{eq:gamma0}
  \gamma_0 \equiv \left\langle 
    \frac{\sum_{i\in \text{charged particles}} \,p_{ti}}%
         {\sum_{i\in \text{all particles}} \,p_{ti}} 
         \right\rangle_\text{\!\!events}\!\!\!\!\!,
\end{equation}
where the sums run over particles in a given event (possibly limited
to some central region with tracking), and the average is carried out
across minimum-bias events.

If the charged pileup particles are not directly included in the
clustering (i.e.\ it is the CHS event that is provided to the
clustering), then one does not have any information on which charged
particles should be used to estimate the neutral pileup in a given
jet.
This problem can be circumvented by a clustering an ``emulated'' CHS
event, in which the charged-pileup particles are kept, but with their
momenta rescaled by an infinitesimal factor $\epsilon$.
In this case the correction becomes
\begin{equation} 
  \label{eq:npc-chs}
  p_\mu^\text{jet,sub} = p_\mu^\text{jet,CHS} -
  \frac{1-\gamma_0}{\gamma_0\, \epsilon}
  p_\mu^\text{jet,rescaled-chg-PU}\,,
\end{equation}
where $p_\mu^\text{jet,CHS}$ is the momentum of the jet as
obtained from the emulated CHS event, while
$p_\mu^\text{jet,rescaled-chg-PU}$ is the summed momentum of the
rescaled charged-pileup particles that are in the jet.
When carrying out NpC-style subtraction, this is our preferred
approach because it eliminates any backreaction associated with the
charged pileup (this is useful also for area-based subtraction), while
retaining the information about charged pileup tracks.

There are multiple issues that may be of concern for the NpC method.
For example, calorimeter fluctuations can limit the experiments'
ability to accurately remove the charged pileup component as measured
with tracks.
For out-of-time pileup, which contributes to calorimetric energy
deposits, charged-track information may not be available at all.
In any case, charged-track information covers only a limited range of
detector pseudorapidities.
Additionally there are subtleties with hadron masses: in effect,
$\gamma_0$ is different for transverse components and for longitudinal
components.
In this work we will avoid this problem by treating all particles as
massless.\footnote{Particle momenta are modified so as to become
  massless while conserving $p_t$, rapidity and azimuth.}
The importance of the above limitations can only be fully evaluated in
an experimental context.

We will be comparing NpC to the area--median method. 
The latter makes a global estimate of pileup
transverse-momentum flow per unit area, $\rho$, by dividing an event
into similarly sized patches and taking the median of the
transverse-momentum per unit area across all the patches.
It then corrects each jet using the globally estimated $\rho$ and the
individual jet's area, $A_\mu$,\footnote{
  With suitable adaptations, the area--median method can be applied to
  characteristics of jets other than the 4-momentum, e.g.\ jet
  shapes~\cite{Soyez:2012hv} and moments of fragmentation
  functions~\cite{Cacciari:2012mu}.  }
\begin{equation} 
  \label{eq:rho-subtraction}
  p_\mu^\text{jet,sub} = p_\mu^\text{jet} -
  \rho A_\mu\,.
\end{equation}
Like NpC, the area--median method has potential experimental
limitations.
They include include questions of non-trivial rapidity dependence and
detector non-linearities (the latter are relevant also for NpC).
These have, to a reasonable extent, been successfully overcome by the
experiments~\cite{ATLAS-PU-Performance,CMS-PU-Performance}.
One respect in which NpC may have advantages over the area--median
method is that the latter fails to correctly account for the fact
that pileup fluctuates from point to point within the event, a feature
that cannot be encoded within the global pileup estimate
$\rho$.\footnote{%
  The area--median $\rho$ determination can be adapted to
  use just the jet's neighbourhood (e.g.\ as discussed in the context
  of heavy-ion collisions~\cite{Cacciari:2010te}), however it can
  never be restricted to just the jet.
}
Furthermore NpC does not need a separate estimation of the background
density $\rho$, which can have systematics related to the event
structure (e.g.\ $t\bar t$ events v.\ dijet events); and there is no
need to include large numbers of ghosts for determining jet areas, a
procedure that has a non-negligible computational cost.

\begin{figure}[t]
  \centering
  \includegraphics[width=0.48\textwidth]{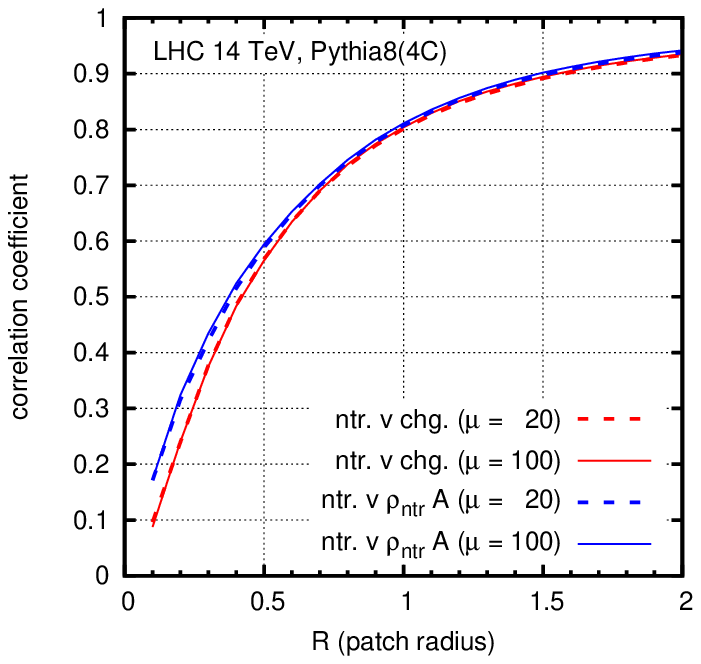}
  \hfill
  \includegraphics[width=0.48\textwidth]{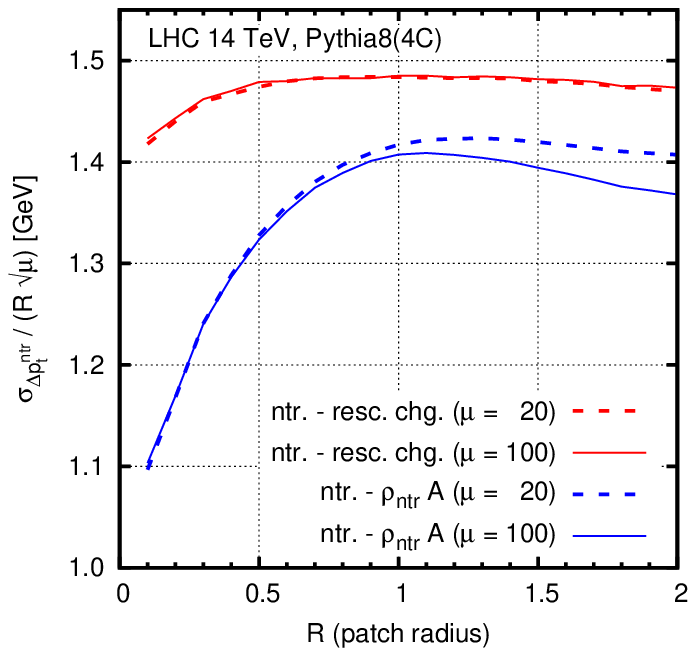}
  \caption{
    Left: the correlation coefficient between the neutral transverse
    momentum in a central patch and either the charged transverse
    momentum in that patch (rescaling this component would not change
    the value of the correlation coefficient) or the prediction using the area--median
    method, i.e.\ $\rho_\text{ntr} A$.
    Right: the standard deviation of the difference between neutral transverse
    momentum in a central patch and either the rescaled charged transverse
    momentum in that patch or the prediction using the area--median
    method, i.e.\ $\rho_\text{ntr} A$.
    The events are composed of superposed zero-bias collisions
    simulated with Pythia~8, tune 4C, and the number of collisions per
    event is Poisson distributed with average $\mu$.}
  \label{fig:correl-central}
\end{figure}

Let us now proceed with an investigation of NpC's performance,
focusing our attention on particle-level events for simplicity.
The key question is the potential performance gain due to NpC's use of
local information.
To study this quantitatively, we consider a circular patch of radius
$R$ centred at $y = \phi = 0$ and examine the correlation coefficient
of the actual neutral energy flow in the patch with two estimates: (a)
one based on the charged energy flow in the same patch and (b) the
other based on a global energy flow determination from the neutral
particles, $\rho_\text{ntr}$.
Fig.~\ref{fig:correl-central} (left) shows these two correlation
coefficients, ``ntr v.\ chg'' and ``ntr v.\ $\rho_\text{ntr} A$'', as
a function of $R$, for two average pileup multiplicities, $\mu = 20$
and $\mu = 100$.
One sees that the local neutral-charged correlation is slightly \emph{lower},
i.e.\ slightly worse, than the neutral-$\rho_\text{ntr}$ correlation.
Both correlations decrease for small patch radii, as is to be
expected, and the difference between them is larger at small
patch radii.
The correlation is largely independent of the number of pileup events
being considered, which is consistent with our expectations, since all
individual terms in the determination of the correlation coefficient
should have the same scaling with $N_\PU$.

Quantitative interpretations of correlation coefficients can sometimes
be delicate, as we discuss in Appendix~\ref{sec:correlation-coefs},
essentially because they combine the covariance of two observables
with the two observables' individual variances.
We find that it can be more robust to investigate a quantity
$\sigma_{\Delta p_t^\text{ntr}}$, the standard deviation of
\begin{equation}
  \label{eq:delta-pt}
  \Delta p_t^\text{ntr} = p_{t}^\text{ntr} - p_{t}^\text{ntr,estimated}\,,
\end{equation}
where the estimate of neutral energy flow,
$p_{t}^\text{ntr,estimated}$, may be either from the rescaled charged
flow or from $\rho_\ntr A$.
The right-hand plot of Fig.~\ref{fig:correl-central} shows
$\sigma_{\Delta p_t^\text{ntr}}$ for the two methods, again as a function of $R$,
for two levels of pileup.
It is normalised to $R \sqrt{\mu}$, to factor out the expected
dependence on both the patch radius and the level of pileup.
A lower value of $\sigma_{\Delta p_t^\text{ntr}}$ implies better performance, and
as with the correlation we reach the conclusion that a global estimate
of $\rho_\ntr$ appears to be slightly more effective at predicting local
neutral energy flow than does the local charged energy flow.
If one hoped to use NpC to improve on the performance of area--median
subtraction, then figure~\ref{fig:correl-central} suggests that
one will be disappointed.

In striving for an understanding of this finding, one should recall
that the ratio of charged-to-neutral energy flow is almost entirely
driven by non-perturbative effects.
Inside an energetic jet, the non-perturbative effects are at scales
$\sim \Lambda_\text{QCD}$ that are tiny compared to the jet transverse
momentum $p_t$.
There are fluctuations in the relative energy carried
by charged and neutral particles, for example because a leading
$u$-quark might pick up a $\bar d$ or a $\bar u$ from the
vacuum. 
However, because $\Lambda_\text{QCD} \ll p_t$, the charged and neutral
energy flow mostly tend to go in the same direction.

The case that we have just seen of an energetic jet gives an intuition
that fluctuations in charged and neutral energy flow are going to be
locally correlated.
It is this intuition that motivates the study of NpC.
We should however examine if this intuition is actually valid for
pileup.
We will examine one step of hadronisation, namely the production of
short-lived hadronic resonances, for example a $\rho^{+}$.
The opening angle between the $\pi^+ \pi^0$ decay products of the
$\rho^{+}$ is of order $2m_\rho / p_{t,\rho}$.
Given that pileup hadrons are produced mostly at low $p_t$, say
$0.5-2\GeV$, and that $m_\rho \simeq 0.77\GeV$, the angle between the
charged and neutral pions ends up being of order $1$ or even larger.
As a result, the correlation in direction between charged and neutral
energy flow is lost, at least in part.
Thus, at low $p_t$, non-perturbative effects specifically tend to wash out the
charged-neutral angular correlation.

\begin{figure}
  \centering
  \begin{minipage}[c]{0.48\linewidth}
    \includegraphics[width=\textwidth]{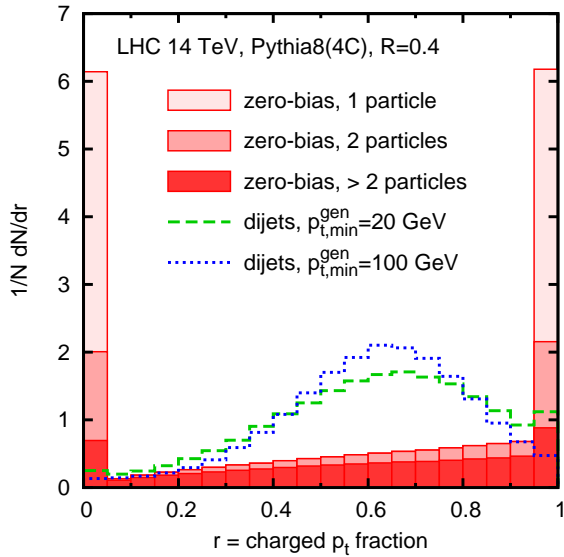}  
  \end{minipage}
  \hfill
  \begin{minipage}[c]{0.48\linewidth}
    \caption{The filled histogram shows the distribution, for simulated
      zero-bias collisions, of the
      fraction, $r$, of the transverse 
      momentum in a central circular patch of radius $R=0.4$ that is due to
      charged particles. 
      It is separated into components according to the multiplicity of
      particles in the patch.
      The dashed and dotted histograms show the corresponding
      charged-fraction distributions for each of the two hardest
      anti-$k_t$, $R=0.4$ jets in simulated dijet events, with two
      choices for the hard generation cut $p_{t,\min}^\text{gen}$.}
    \label{fig:why-NpC-bad}
  \end{minipage}
\end{figure}

This point is illustrated in Fig.~\ref{fig:why-NpC-bad}.
We consider zero-bias events and examine a circular patch of radius
$R=0.4$ centred at $y = \phi = 0$. 
The figure shows the distribution of the charged $p_t$ fraction, $r$,
\begin{equation}
  \label{eq:r-definition}
  r = \frac{p_{t}^\text{chg}}{p_{t}^{\text{chg}+\text{ntr}}}\,,
\end{equation}
in the patch (filled histogram, broken into contributions where the
patch contains $1$, $2$ or more particles).
The same plot also shows the distribution of the charged $p_t$ fraction
in each of the two leading anti-$k_t$, $R=0.4$ jets in dijet events
(dashed and dotted histograms). 
Whereas the charged-to-total ratio for a jet has a distribution peaked
around $0.6$, as one would expect, albeit with a broad distribution,
the result for zero-bias events is striking: in about 60\% of events the
patch is either just charged or just neutral, 
quite often consisting of just a single particle (weighting by the $p_t$ 
flow in the patch, the figure goes down to 30\%).
This is probably part of the reason why charged information provides
only limited local information about neutral energy flow in pileup
events.

These considerations are confirmed by an analysis of the actual
performance of NpC and area--median subtraction.
We reconstruct jets using the anti-$k_t$
algorithm~\cite{Cacciari:2008gp}, as implemented in
FastJet\footnote{Results shown in this paper have been obtained in
  some cases with a development snapshot of version 3.1, in others
  with versions 3.1.0 and 3.1.1.}~\cite{FastJet}, with a jet radius
parameter of $R=0.4$.
We study dijet and pileup events generated with
Pythia~8.176~\cite{Pythia8}, in tune 4C; we assume idealised CHS,
treating the charged pileup particles as ghosts.
In the dijet (``hard'') event alone, i.e.\ without pileup, we run the
jet algorithm and identify jets with absolute rapidity $|y|<2.5$ and
transverse momentum $p_t> 150\GeV$.
Then in the event with superposed pileup (the ``full'' event) we rerun
the jet algorithm and identify the jets that match those selected in
the hard event\footnote{\label{footnote:matching}
  For the matching, we introduce a quantity
  $p_t^\text{shared}(j_i^\text{hard}, j_j^\text{full})$, the scalar sum
  of the $p_t$'s of the constituents that are common to a given pair $i,j$ of hard and
  full jets.
  For a hard jet $i$, the matched jet in the full event is the one that
  has the largest $p_t^\text{shared}(j_i^\text{hard},
  j_j^\text{full})$. 
  In principle, one full jet can match two hard jets, e.g.\ if two
  nearby hard jets end up merged into a single full jet due to
  back-reaction effects. However this is exceedingly rare.
}
and subtract them using either NpC,
Eq.~(\ref{eq:npc-chs}), or the area--median method,
Eq.~(\ref{eq:rho-subtraction}), with $\rho$ estimated from the CHS
event.
The hard events are generated with the underlying event turned off, which
enables us to avoid subtleties related to the simultaneous subtraction
of the underlying event.

\begin{figure}
  \centering
  \includegraphics[width=0.48\textwidth]{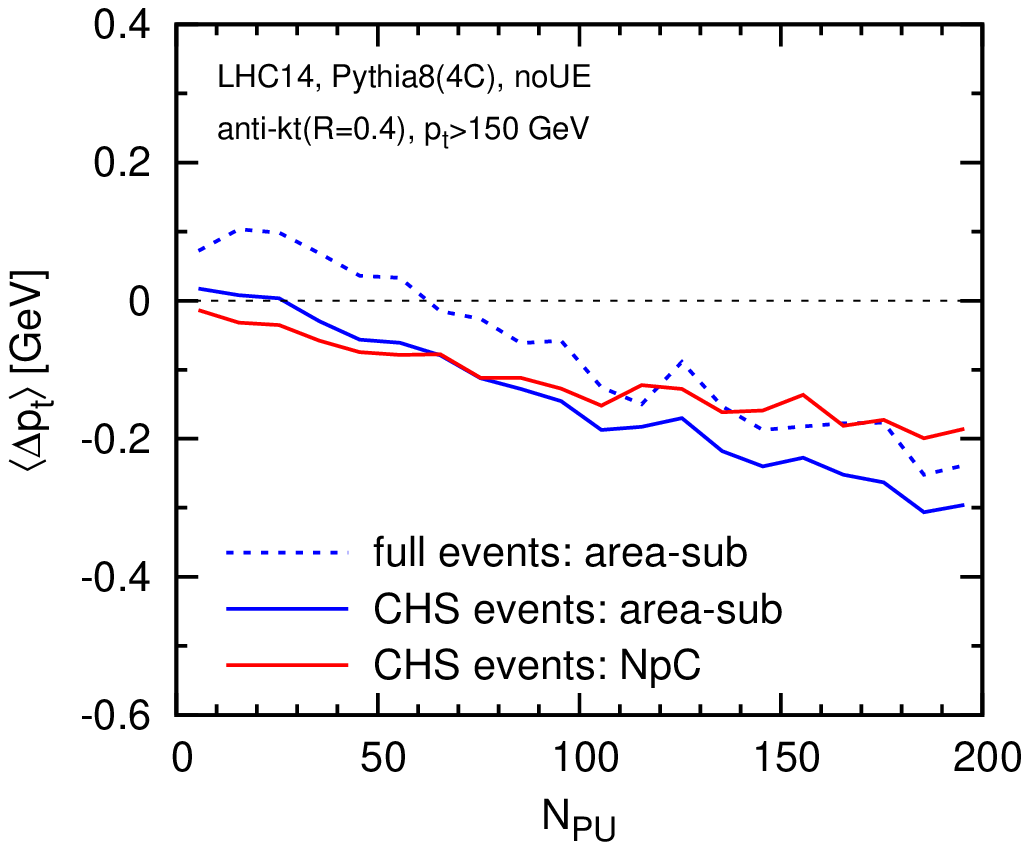}
  \hfill
  \includegraphics[width=0.48\textwidth]{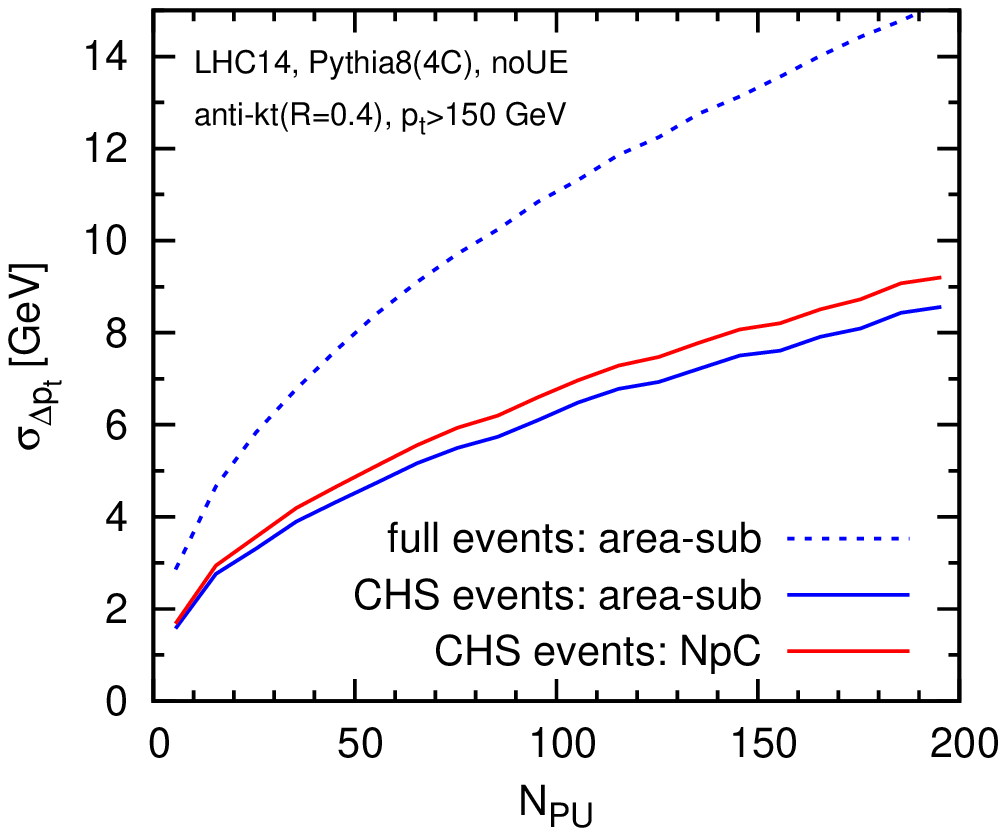}
  \caption{A comparison of the performance of the NpC and area--median 
    subtraction methods. 
    The left-hand plot shows, as a function of the
    number of pileup vertices $N_\text{PU}$, the average
    difference in $p_t$
    between a jet after pileup addition and subtraction and the
    corresponding matched jet in the hard sample, $\Delta p_t=p_t^{\rm
      jet,sub}-p_t^{\rm jet,hard}$. 
    The right-hand plot shows the standard deviation of $\Delta p_t$
    (lower values are better).
    NpC is shown only for CHS events, while area--median subtraction
    is shown both for events with CHS and for events without it
    (``full''). }
  \label{fig:npc-performance}
\end{figure}

Figure~\ref{fig:npc-performance} provides the resulting comparison of
the performance of the NpC and area--median subtraction methods (the
latter in CHS and in full events).
The left-hand plot shows the average difference between the subtracted
jet $p_t$ and the $p_t$ of the corresponding matched hard jet, as a
function of the number of pileup interactions.
Both methods clearly perform well here, with the average difference
systematically well below $1\GeV$ even for very high pileup levels.
The right-hand plot shows the standard deviation of the difference
between the hard and subtracted full jet $p_t$. 
A lower value indicates better performance, and one sees that in CHS
events the area--median method indeed appears to have a small, but
consistent advantage over NpC.
Comparing area--median subtraction in CHS and full events, one
observes a significant degradation in resolution when one fails to use
the available information about charged particles in correcting the
charged component of pileup, as is to be expected for a particle-level
study.

The conclusion of this section is that the NpC method fails to give a
superior performance to the area--median method in CHS events. This is
because the local correlations of neutral and charged energy flow are
no greater than the correlations between local neutral energy flow and
the global energy flow.
We believe that part of the reason for this is that the hadronisation
process for low $p_t$ particles intrinsically tends to produce hadrons
separated by large angles, as illustrated concretely in the case of
$\rho^\pm$ resonance decay.

%======================================================================
\section{Cleansing}
\label{sec:appraisal}

Part of the original motivation for our work here was to cross check a
method recently introduced by Krohn, Low, Schwartz and Wang (KLSW) and
called jet cleansing~\cite{Krohn:2013lba}.
Cleansing comes in several variants.
We will concentrate on linear cleansing, which was seen to
perform well across a variety of observables by KLSW.\footnote{Other
  variants of cleansing introduced by KLSW 
  include ``Jet-Vertex-Fraction'' (JVF) and ``Gaussian'' versions.
  JVF scales each subjet by
  $p_{t}^\text{chg-LV}/p_{t}^\text{chg-total}$, the ratio of the
  charged $p_t$ from the leading vertex to the total charged $p_t$
  (including pileup) in the subjet.
  Gaussian is particularly interesting in that it effectively carries
  out a $\chi^2$ minimisation across different hypotheses for the ratio
  of charged to neutral energy flow, separately for the pileup and the
  hard event. However in KLSW's results its performance was usually
  only marginally better than the much simpler linear cleansing. 
  Accordingly we concentrate on the latter.  }
It involves several elements: it breaks a jet into multiple subjets,
as done for grooming methods like filtering and trimming~\cite{trim,
  filter} (cf.\ also the early work by Seymour~\cite{seymour}).
In its ``linear'' variant, it then corrects individual subjets for
pileup by a method that is essentially the same as the NpC approach
described in the previous section.
Cleansing may also be used in conjunction with trimming-style cuts to
the subtracted subjets, specifically it can remove those whose
corrected transverse momentum is less than some fraction
$f_\text{cut}$ of the overall jet's transverse momentum (as evaluated
before pileup removal).\footnote{%
  In v1 of this article as submitted to arXiv in April 2014, and also
  in a version circulated to the authors of Ref.~\cite{Krohn:2013lba}
  several months before the arXiv submission, we used
  $f_\text{cut}=0.05$ for cleansing, reflecting our understanding of
  the choices made in v1 of Ref.~\cite{Krohn:2013lba}, which stated
  ``[we] supplement cleansing by applying a cut on the ratio $f$ of
  the subjet $p_T$ (after cleansing) to the total jet $p_T$. Subjets
  with $f < f_\cut$ are discarded. [...]  Where we do trim/cleanse we
  employ $R_\text{sub}=0.3$ subjets and take $f_\text{cut}=0.05$.''
  Subsequent to the appearance of v1 of our article, the authors of
  Ref.~\cite{Krohn:2013lba} clarified that the results in their Fig.~4
  had used $f_\cut=0$.
  This is the choice that we adopt throughout most of this version,
  and it has an impact notably on the conclusions for the jet-mass
  performance.}

\begin{figure}[pt]
  \centering
  \begin{minipage}{0.48\linewidth}
    \includegraphics[width=\textwidth]{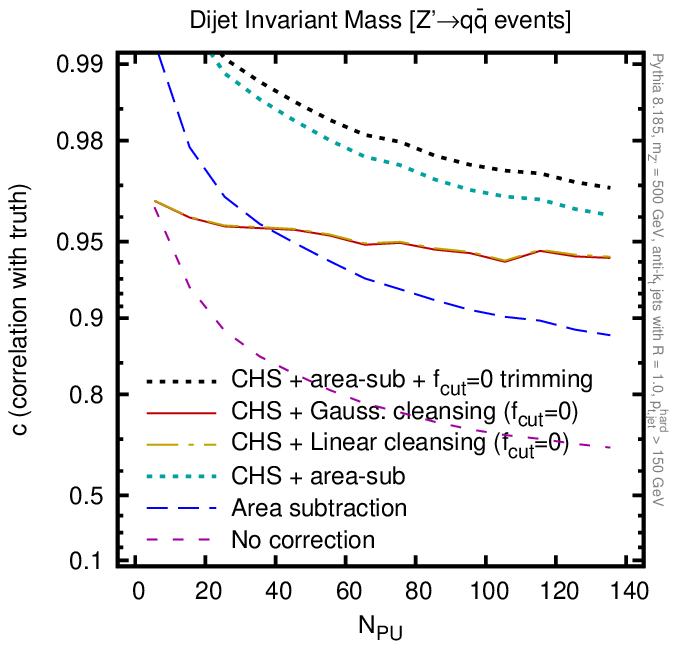}
  \end{minipage}
  \hfill
  \begin{minipage}{0.48\linewidth}
    \includegraphics[width=\textwidth]{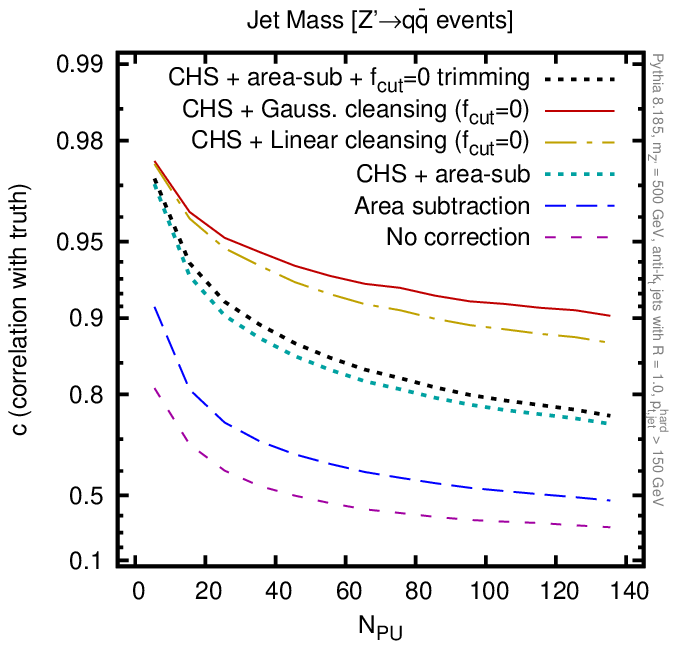}%
  \end{minipage}%
  \\
  \begin{minipage}{0.48\linewidth}
    \includegraphics[width=\textwidth]{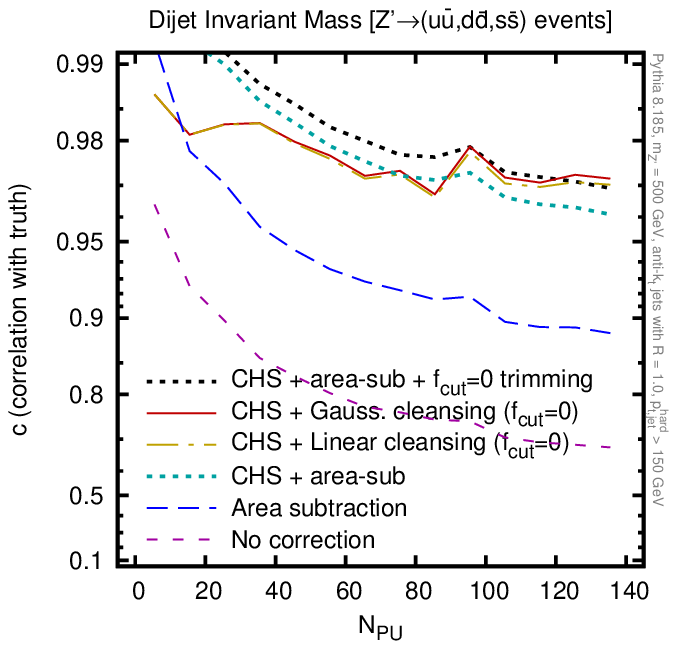}
  \end{minipage}
  \hfill
  \begin{minipage}{0.48\linewidth}
    \caption{
        Upper left plot: the correlation coefficient between the dijet
        mass in hard events and the dijet mass after adding pileup and
        applying various pileup-removal methods; shown as a function of
        the number of pileup events, $N_\text{PU}$.
        Upper right plot: similarly for the single jet mass.
        Both plots are for a hadronically decaying $Z'$ sample with $m_{Z'} =
        500\GeV$.
        Decays are included to all flavours except $t\bar t$ and
        $B$-hadrons are taken stable. 
        Lower-left plot: the correlation coefficient for the dijet-mass, as
        in the upper-left plot, but with a sample of $Z'$ bosons that decay
        only to $u$, $d$ and $s$ quarks.
        Jets are reconstructed, as in Ref.~\cite{Krohn:2013lba}, with
        the anti-$k_t$ algorithm with $R=1$. 
        For both trimming and cleansing, subjets are reconstructed with
        the $k_t$ algorithm with $R_\text{sub}=0.3$ and the
        $f_\cut$ value that is applied is $f_\cut=0$.
      %}
      \label{fig:f00}
    }
  \end{minipage}
\end{figure}

The top left-hand plot of Fig.~\ref{fig:f00} shows the correlation
coefficient between the dijet mass in a hard event and the dijet
mass after addition of pileup and application of each of several pileup
mitigation methods.
The results are shown as a function of $N_\PU$.
The pileup mitigation methods include two forms of cleansing (with
$f_\cut=0$), 
area--median subtraction, 
CHS+area subtraction, and CHS+area subtraction in
conjunction with trimming (also with $f_\cut=0$).
The top right-hand plots shows the corresponding results for the jet mass.
For the dijet mass we 
see that linear (and Gaussian)
cleansing performs worse than area subtraction, while in the
right-hand plot, for the jet mass, we 
see linear (and Gaussian)
cleansing performing better than area subtraction, albeit not to the
extent found in Ref.~\cite{Krohn:2013lba}.
These (and, unless explicitly stated, our other $Z'$ results) have
been generated with the $Z'$ decaying to all flavours except
$t\bar t$, and $B$-hadrons have been kept stable.\footnote{%
  We often find this to be useful for particle-level $b$-tagging
  studies.
  Experimentally, in the future, one might even imagine an
  ``idealised'' form of particle flow that attempts to reconstruct
  $B$-hadrons (or at least their charged part) from displaced tracks
  before jet clustering. }
The lower plot shows the dijet mass for a different $Z'$ sample, one
that decays only to $u$, $d$ and $s$ quarks, but not $c$ and $b$
quarks. 
Most of the results are essentially unchanged. 
The exception is cleansing, which turns out to be very sensitive to
the sample choice.
Without stable $B$-hadrons in the sample, its performance improves
noticeably and at high pileup becomes comparable to that of
area-subtraction.
Both of the left-hand plots in our Fig.~\ref{fig:f00} differ
noticeably from Fig.~4 (left) of Ref.~\cite{Krohn:2013lba} and in
particular they are not consistent with KLSW's observation of much
improved correlation coefficients for the dijet mass with cleansing
relative to area+CHS subtraction.%
\footnote{%
  We remain puzzled also by the relative pattern of area+CHS v.\
  area-subtracted results in Fig.~4 (left) of
  Ref.~\cite{Krohn:2013lba}, since the area+CHS curves appears to tend
  towards area at large pileup, whereas the use of CHS should 
  significantly reduce the impact of pileup.
}

Given our results on NpC in section~\ref{sec:NpC}, we were puzzled by the difference between
the performance of area-subtraction plus trimming versus that of
cleansing: our expectation is that their performances should be very
similar.\footnote{KLSW state in \cite{Krohn:2013lba} that fluctuations
  around a `best' charged fraction $\bar \gamma_0$ decrease with
  increasing $N_\PU$ and suggest (see also
  \cite{schwartz-talk-PU-workshop}, pp. 16 and 17) that this will
  improve the determination of this fraction and therefore the
  effectiveness of a method like cleansing, based on a
  neutral-proportional-to-charged approach.  
  However, this does not happen because, while relative fluctuations
  around $\bar \gamma_0$ do indeed decrease proportionally to
  $1/\sqrt{N_\PU}$ (a result of the incoherent addition of many pileup
  events and of the Central Limit Theorem), the absolute uncertainty
  that they induce on a pileup-subtracted quantity involves an
  additional factor $N_\PU$.
  The product of the two terms is therefore proportional to
  $\sqrt{N_\PU}$, i.e. the same scaling as the area-median
  method. 
  This is consistent with our observations.
  Note that for area subtraction, the switch from full events to CHS
  events has the effect of reducing the coefficient in front of
  $\sqrt{N_\PU}$.  }
The strong sample-dependence of the cleansing performance also calls
for an explanation.
We thus continued our study of the question.

According to the description in Ref.~\cite{Krohn:2013lba}, one
additional characteristic of linear cleansing relative to
area-subtraction is that it switches to jet-vertex-fraction (JVF)
cleansing when the NpC-style rescaling would give a negative answer.
In contrast, area-subtraction plus trimming simply sets the (sub)jet
momentum to zero.
We explicitly tried turning the switch to JVF-cleansing on and off and
found it had a small effect and did not explain the differences.

Study of the public code for jet cleansing\footnote{Version 1.0.1 from
  fjcontrib~\cite{fjcontrib}.} reveals an additional
condition being applied to subjets: if a subjet contains no charged
particles from the leading vertex (LV), then its momentum is set to
zero.
This step appears not to have been mentioned in
Ref.~\cite{Krohn:2013lba}.
Since we will be discussing it extensively, we find it useful to give
it a name, ``\emph{zeroing}''.
Zeroing can be thought of as an extreme limit of the charged-track
based trimming procedure introduced by ATLAS~\cite{ATL-Pileup-2},
whereby a JVF-style cut is applied to reject subjets whose
charged-momentum fraction from the leading vertex is too
low.
Zeroing turns out to be crucial: if we use it in conjunction with CHS
area-subtraction (or with NpC subtraction) and $f_\cut=0$ trimming, we
obtain results that are very similar to those from cleansing.
Conversely, if we turn this step off in linear-cleansing, its results come into
accord with those from (CHS) area-subtraction or NpC-subtraction with
$f_\cut=0$ trimming.

\begin{figure}[tp]
  \centering
  \includegraphics[width=0.48\textwidth]{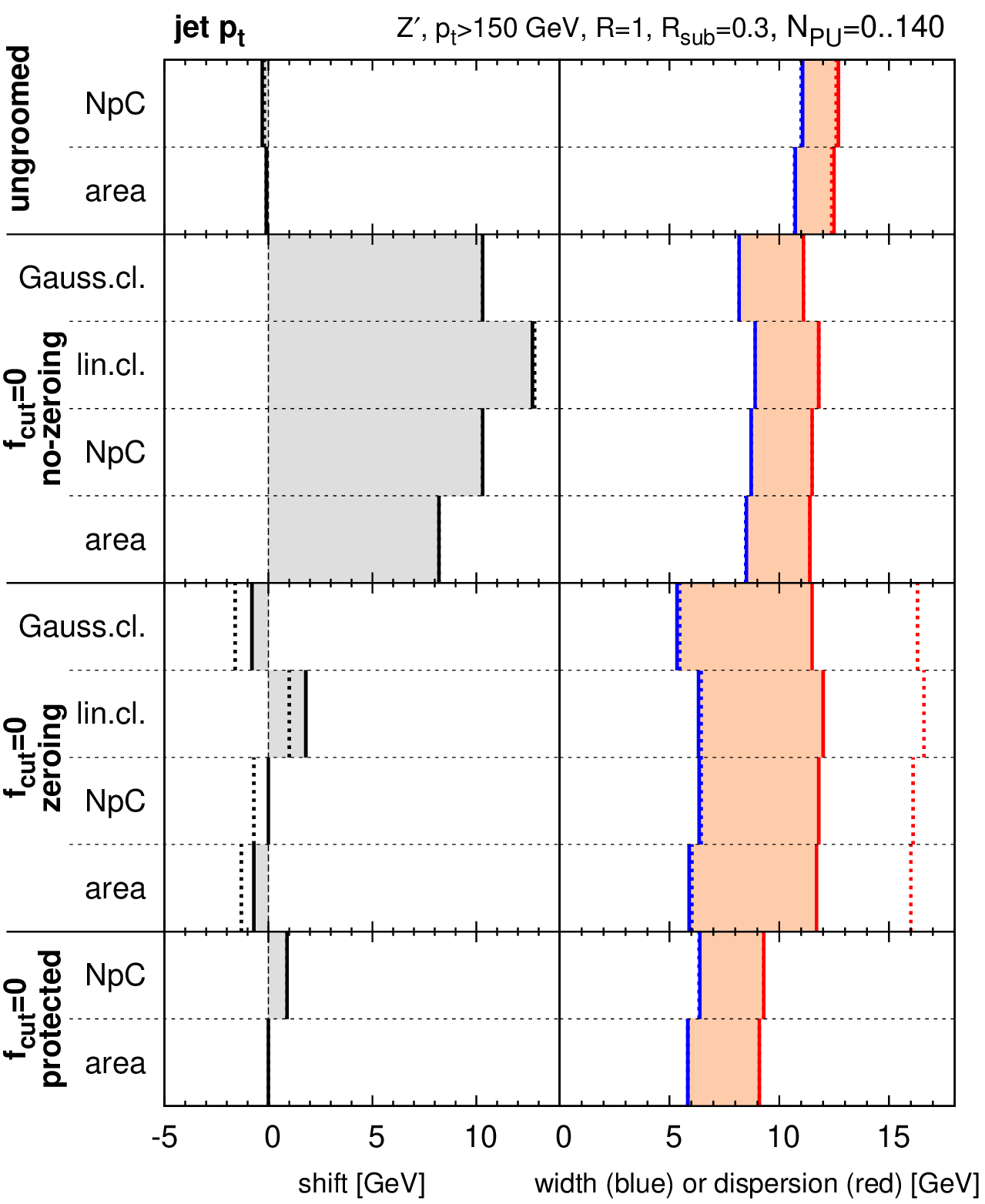}
  \hfill
  \includegraphics[width=0.48\textwidth]{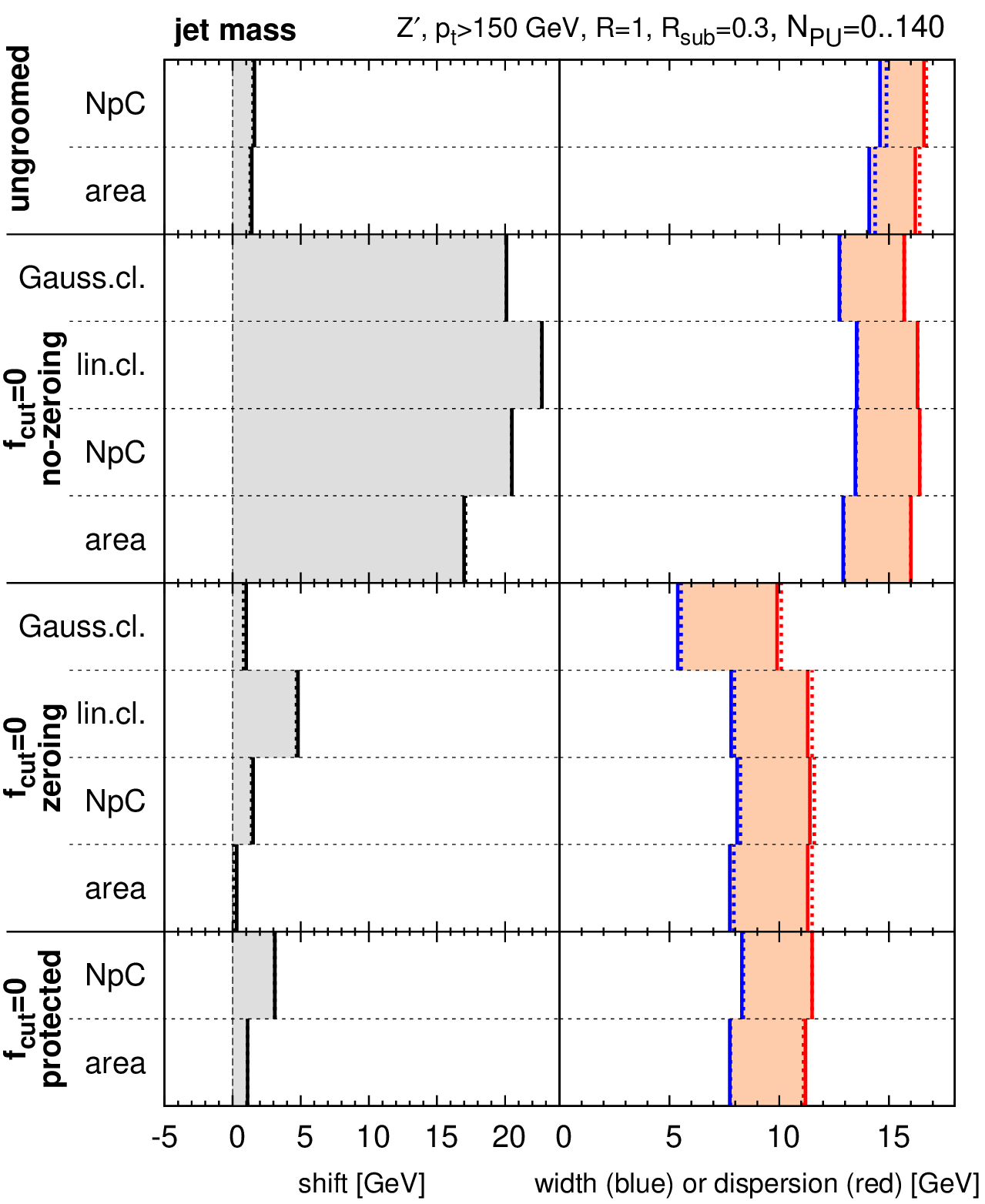}
  \caption{The left-hand plot illustrates various characteristics of the
    change ($\Delta p_t$) in the jet $p_t$ after addition of pileup and removal
    by a range of methods.
    It shows the average shift $\langle \Delta p_t
    \rangle$ (in black) and the peak width (in blue) and dispersion
    (in red) of the $\Delta p_t$ distribution. 
    The peak width is defined as the smallest window of $\Delta p_t$ that
    contains 90\% of the $\Delta p_t$ distribution, scaled by a
    factor $\simeq 0.304$
    such that in the case of a Gaussian distribution the result agrees
    with the dispersion.
    The right-hand plot shows the same set of results for the jet mass.
    The results are obtained in a sample of events with the number of
    pileup vertices distributed uniformly between $0$ and $140$.
    The hard events consist of hadronic $Z'$ decays: for the solid
    vertical lines the sample is $Z'\to d\bar d, u\bar u, s\bar s$,
    while for the dotted lines (sometimes not visible because directly
    over the solid lines), the sample
    additionally includes $Z'\to
    c\bar c, b\bar b$ with $B$ hadrons kept stable.
    The $Z'$ mass is $m_{Z'}=500\GeV$
    and jets are reconstructed with the anti-$k_t$ algorithm with $R=1$.
    All results in this figure include charged-hadron subtraction by default.
    The default form of cleansing, as used e.g.\ in Fig.~\ref{fig:f00},
    is ``$f_\cut=0$ zeroing''.  
  }
  \label{fig:shifts-dispersions-R1}
\end{figure}

To help illustrate this, Fig.~\ref{fig:shifts-dispersions-R1} shows a
``fingerprint'' for each of several pileup-removal methods, for both
the jet $p_t$ (left) and mass (right).
The fingerprint includes the average shift
($\langle \Delta p_t\rangle$ or $\langle \Delta m\rangle$) of the
observable after pileup removal, shown in black.
It also includes two measures of the width of the $\Delta p_t$ and
$\Delta m$ distributions: the dispersion (i.e. standard deviation) in
red and an alternative peak-width measure in blue. 
The latter 
is defined as follows: one determines the
width of the smallest window that contains $90\%$ of entries and then
scales this width by a factor $0.304$.
For a Gaussian distribution, the rescaling ensures that the resulting
peak-width measure is equal to the dispersion.
For a non-Gaussian distribution the two measures usually differ and
the orange shaded region quantifies the extent of this difference. 
The solid black, blue and red lines have been obtained from samples in
which the $Z'$ decays just to light quarks; the dotted lines are for a
sample including $c\bar c$ and $b\bar b$ decays (with stable
$B$-hadrons), providing an indication of the sample dependence; in
many cases they are indistinguishable from the solid lines.

Comparing $f_\cut=0$ grooming for NpC, area (without zeroing) and cleansing
with zeroing manually disabled, all have very similar fingerprints.
Turning on zeroing in the different methods leads to a significant
change in the fingerprints, but again NpC, area and cleansing are very
similar.\footnote{For the jet mass, Gaussian cleansing appears a
  little different in the $f_\cut=0$ case with zeroing, suggesting that
  there may be an advantage from combinations of different constraints
  on subjet momenta.
}

When used with $f_\cut=0$ trimming, and when examining quality
measures such as the dispersion (in red, or the closely related correlation
coefficient, cf.\ Appendix~\ref{sec:correlation-coefs}), subjet
zeroing appears to be advantageous for the jet mass, but potentially
problematic for the jet $p_t$ and the dijet mass.
However, the dispersion quality measure does not tell the full story
regarding the impact of zeroing.
Examining simultaneously the peak-width measure (in blue) makes it
easier to disentangle two different effects of zeroing.
On one hand we find that zeroing correctly rejects subjets that are
entirely due to fluctuations of the pileup.
This narrows the peak of the $\Delta p_t$ or $\Delta m$ distribution,
substantially reducing the (blue) peak-width measures in
Fig.~\ref{fig:shifts-dispersions-R1}.
On the other hand, zeroing sometimes incorrectly rejects subjets that
have no charged tracks from the LV but do have significant
neutral energy flow from the LV.
This can lead to long tails for the $\Delta p_t$ or $\Delta m$
distributions, adversely affecting the dispersion.\footnote{A
  discrepancy between dispersion and peak-width measures is to be seen
  in Fig.~15 of Ref.~\cite{CMS:2014ata} for jet masses. Our
  ``fingerprint'' plot is in part inspired by the representation
  provided there, though our choice of peak-width measure differs.}
It is the interplay between the narrower peak and the longer tails
that affects whether overall the dispersion goes up or down with
zeroing.
In particular the tails appear to matter more for the jet $p_t$ and
dijet mass than they do for the single-jet mass.
Note that accurate Monte Carlo simulation of such tails may be quite
challenging: they appear to be associated with configurations where a
subjet contains an unusually small number of energetic neutral
particles. 
Such configurations are similar to those that give rise to fake
isolated photons or leptons and that are widely known to be difficult
to simulate correctly.

We commented earlier that the cleansing performance has a significant
sample dependence.
This is directly related to the zeroing: indeed
Fig.~\ref{fig:shifts-dispersions-R1} shows that for cleansing without
zeroing, the sample dependence (dashed versus solid lines) vanishes,
while it is substantial with zeroing.
Our understanding of this feature is that the lower multiplicity of
jets with undecayed $B$-hadrons (and related hard fragmentation of the
$B$-hadron) results in a higher likelihood that a subjet will contain
neutral but no charged particles from the LV, thus enhancing the
impact of zeroing on the tail of the $\Delta p_t$ or $\Delta m_{jj}$
sample.

The long tails produced by the zeroing are not necessarily
unavoidable.
In particular, they can correspond to the loss of subjets with tens of
GeV, yet it is very unlikely that a subjet from a pileup collision
will be responsible for such a large energy.
Therefore we introduce a modified procedure that we call
``\emph{protected zeroing}'': one rejects any subjet without LV tracks
\emph{unless} its $p_t$ after subtraction is $n$ times larger than the
largest charged $p_t$ in the subjet from any single pileup vertex (or,
more simply, just above some threshold $p_{t,\text{min}}$; however,
using $n$ times the largest charged subjet $p_t$ could arguably be
better both in cases where one explores a wide range of $N_\PU$ and
for situations involving a hard subjet from a pileup collision).
Taking $n=10$ (or a fixed $p_{t,\text{min}} = 20\GeV$) we have found
reduced tails and, consequently, 
noticeable improvements in the jet $p_t$ and dijet mass dispersion
(with little effect for the jet mass).
This is visible for area and NpC subtraction in
Fig.~\ref{fig:shifts-dispersions-R1}.
Protected zeroing also eliminates the sample dependence.%
\footnote{%
  We learned while finalizing this note that for the identification of
  pileup v.\ non-pileup full jets in ATLAS, some form of protection is
  already in place, in that JVF-type conditions are not applied if
  $p_t > 50 \GeV$.  We thank David Miller for exchanges on this point.
}

Several additional comments can be made about $f_\cut=0$ trimming combined with
zeroing.
Firstly, $f_\cut=0$ trimming alone introduces a bias in the jet
$p_t$, which is clearly visible in the $f_\cut=0$ no-zeroing shifts in
Fig.~\ref{fig:shifts-dispersions-R1}. 
This is because the trimming removes negative fluctuations of the
pileup, but keeps the positive fluctuations.
Zeroing then counteracts that bias by removing some of the positive
fluctuations, those that happened not to have any charged tracks from
the LV.
It also introduces further negative fluctuations for subjets that
happened to have some neutral energy flow but no charged tracks.
Overall, one sees that the final net bias comes out to be relatively
small.
This kind of cancellation between different biases is common in
noise-reducing pileup-reduction
approaches~\cite{Kodolova,Cacciari:2014gra,Bertolini:2014bba}.

\begin{figure}[tp]
  \centering
  \begin{minipage}{0.48\linewidth}
    \includegraphics[width=\textwidth]{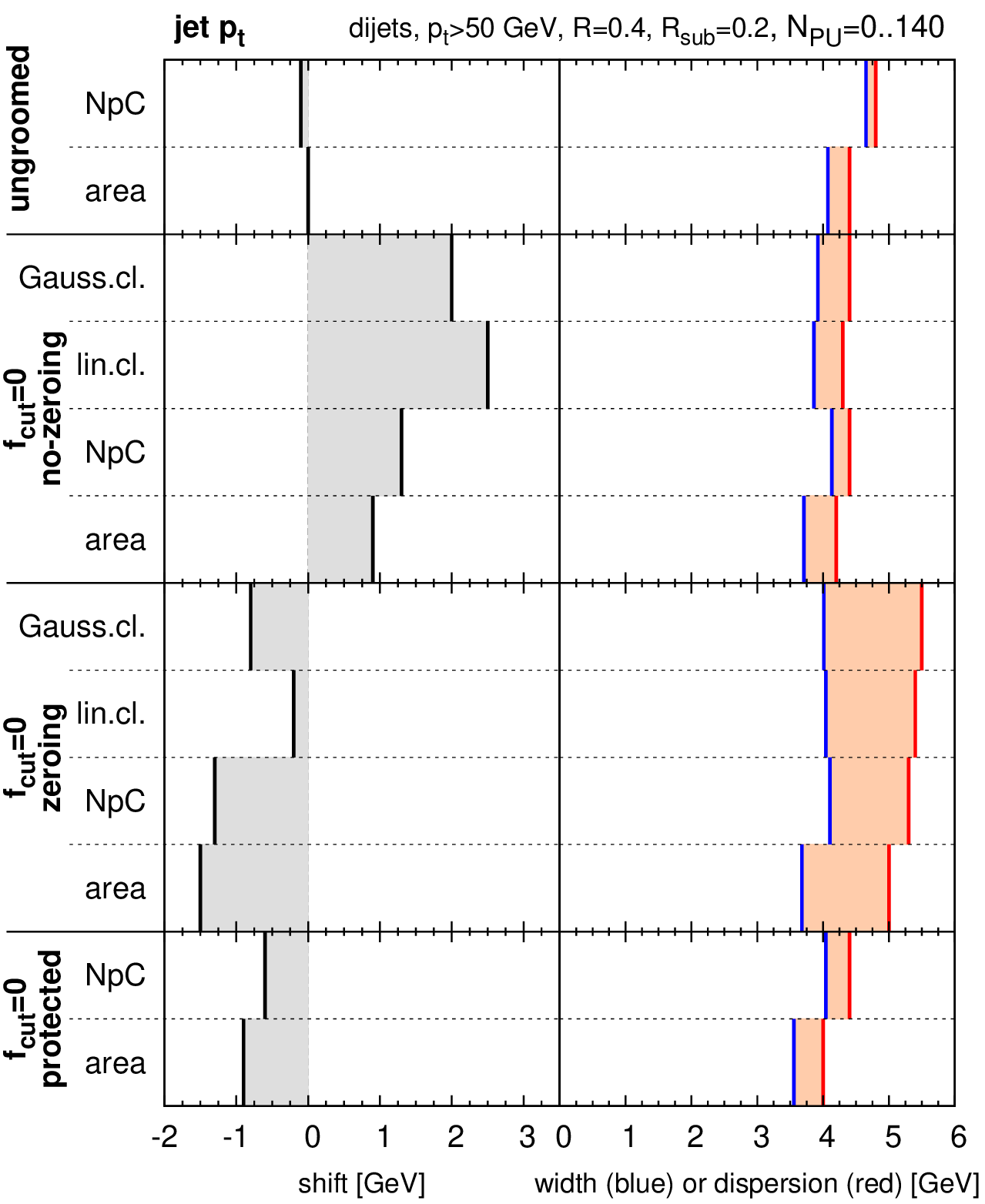}
  \end{minipage}
  \hfill
  \begin{minipage}{0.48\linewidth}
    \includegraphics[width=\textwidth]{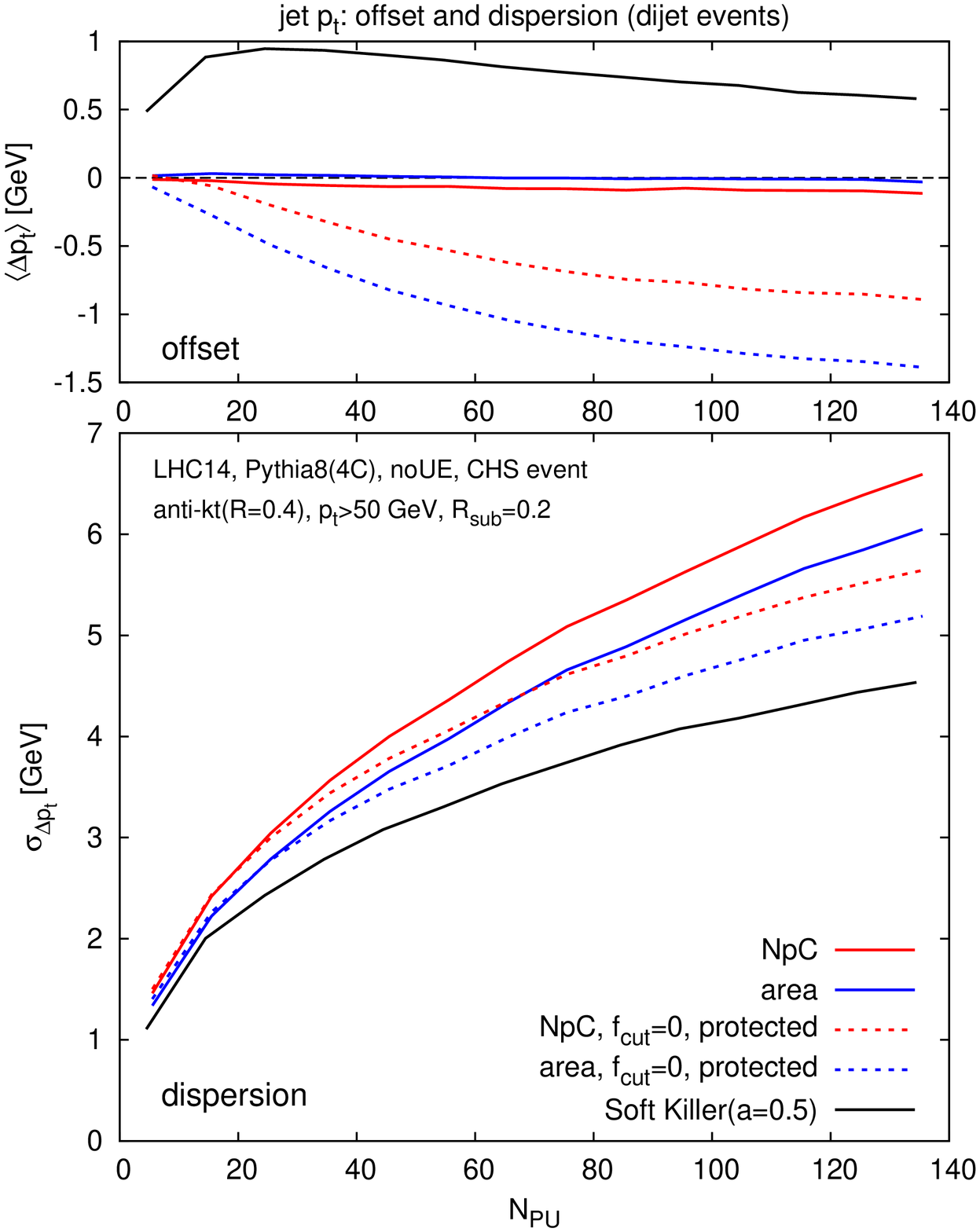}
  \end{minipage}
  %
  %\begin{minipage}{1.0\linewidth}
    \caption{
      Left, Analogue of Fig.~\ref{fig:shifts-dispersions-R1} (left) for a
      jet radius of $R=0.4$, subjet radius (where relevant) of
      $R_\text{sub}=0.2$ and a QCD continuum dijet sample generated with Pythia~8.
      The underlying event is turned off in the sample and $B$ hadrons
      decay. 
      We consider only jets that in the hard sample have $p_t > 50\GeV$
      and $|y|< 2.5$.
      Right: the dispersions for a subset of the methods, shown as a function
      of the number of pileup events.
      \label{fig:shifts-dispersions-dijet-R0.4}
    }
  %\end{minipage}
\end{figure}

\begin{figure}[tp]
  \centering
  \begin{minipage}{0.48\linewidth}
    \includegraphics[width=\textwidth]{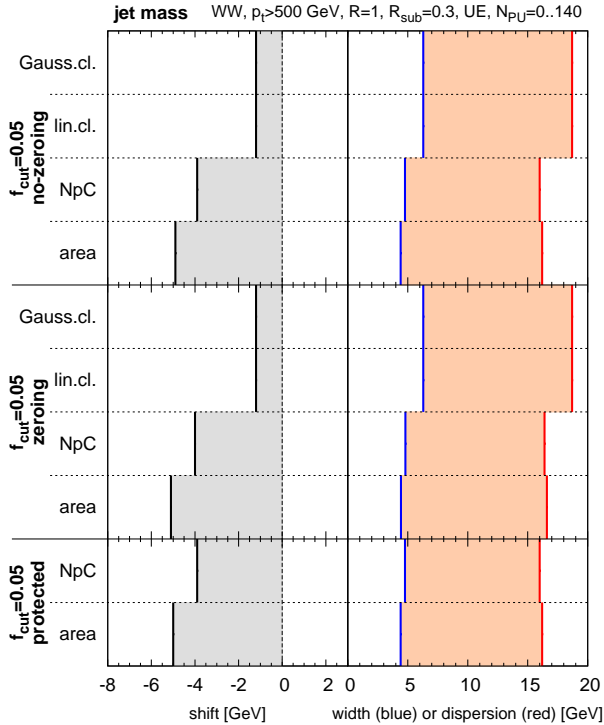}
  \end{minipage}
  \hfill
  \begin{minipage}{0.48\linewidth}
    \caption{
      Analogue of Fig.~\ref{fig:shifts-dispersions-R1} (right),
      showing the performance for the jet mass, but
      now with $f_\cut=0.05$ applied to both trimming and cleansing and in
      a sample of hadronically-decaying boosted $W$ bosons ($pp \to
      W^+W^-$).
      The jets reconstructed after addition and subtraction of pileup
      are compared to trimmed hard jets.
      Jets are reconstructed with a jet radius of $R=1$ and a subjet
      radius of $R_\text{sub} = 0.3$. 
      Only hard jets with $p_t > 500\GeV$ and $|y| < 2.5$ (before
      trimming) are considered and we let $B$ hadrons decay.
      \label{fig:shifts-dispersions-WW500-R10}
    }
  \end{minipage}
\end{figure}

Most of the studies so far in this section have been carried out with a setup
that is similar to that of Ref.~\cite{Krohn:2013lba}, i.e.\ $R=1$ jets
in a $Z'$ sample with $f_\cut=0$ trimming.
This is not a common setup for most practical applications.
For most uses of jets, $R=0.4$ is a standard choice and pileup is
at its most severe at low to moderate $p_t$.
Accordingly, in Fig.~\ref{fig:shifts-dispersions-dijet-R0.4} (left) we show
the analogue of Fig.~\ref{fig:shifts-dispersions-R1}'s summary for the
jet $p_t$, but now for $R=0.4$, with $R_\text{sub}=0.2$ in a QCD dijet
sample, considering jets that in the hard event had $p_t > 50\GeV$.
We see that qualitatively the pattern is quite similar to that in
Fig.~\ref{fig:shifts-dispersions-R1}.\footnote{Note, however, that at
  even lower jet $p_t$'s, the difference between zeroing and protected
  zeroing might be expected to disappear. This is because the long
  negative tails are suppressed by the low jet $p_t$ itself.}
Quantitatively, the difference between the various choices is much
smaller, with about a $10\%$ reduction in dispersion (or width) in going
from ungroomed CHS area-subtraction to the $f_\cut=0$ protected
subjet-zeroing case.
One should be aware that this study is only for a single $p_t$, across
a broad range of pileup.
The dispersions for a subset of the methods are shown as a function of
the number of pileup vertices in the right-hand plot of
Fig.~\ref{fig:shifts-dispersions-dijet-R0.4}.
That plot also includes results from the SoftKiller
method~\cite{Cacciari:2014gra} and illustrates that the benefit from
protected zeroing (comparing the solid and dashed blue curves) is
about half of the benefit that is brought from SoftKiller (comparing
solid blue and black curves).
These plots show that protected zeroing is potentially of interest for
jet $p_t$ determinations in realistic conditions. 
Thus it would probably benefit from further study: one should, for
example, check its behaviour across a range of transverse momenta,
determine optimal choices for the protection of the zeroing and
investigate also how best to combine it with particle-level
subtraction methods such as SoftKiller.\footnote{An interesting
  feature of protected zeroing, SoftKiller and another
  recently introduced method, PUPPI~\cite{Bertolini:2014bba}, is that
  the residual degradation in resolution from pileup appears to scale
  more slowly than the $\sqrt{N_\PU}$ pattern that is observed for
  area and NpC subtraction alone.}

Turning now to jet masses, the use of $R=1$ is a not uncommon choice,
however most applications use a groomed jet mass with a non-zero
$f_\cut$ (or its equivalent): this improves mass resolution in the
hard event even without pileup, and it also reduces backgrounds,
changing the perturbative structure of the
jet~\cite{Dasgupta:2013ihk,Dasgupta:2013via} even in the absence of
pileup.\footnote{In contrast, for $f_\cut=0$ trimming, the jet
  structure is unchanced in the absence of pileup.}
Accordingly in Fig.~\ref{fig:shifts-dispersions-WW500-R10} we show
$f_\cut=0.05$ results (with shifts and widths computed relative to
$f_\cut=0.05$ trimmed hard jets) for a hard $WW$ sample where the hard
fat jets are required to have $p_t > 500\GeV$. 
Zeroing, whether protected or not, appears to have little impact. 
One potential explanation for this fact is as follows: zeroing's
benefit comes primarily because it rejects fairly low-$p_t$ pileup
subjets that happen to have no charged particles from the leading
vertex.
However for a pileup subjet to pass the $f_\cut=0.05$ filtering
criterion in our sample, it would have to have $p_t > 25\GeV$.
This is quite rare.
Thus filtering is already removing the pileup subjets, with little
further to be gained from the charged-based zeroing.
As in the plain jet-mass summary plot, protection of zeroing appears
to have little impact for the trimmed jet mass.\footnote{
  \label{footnote:trimming-ref}
  Cleansing
  appears to perform slightly worse than trimming with NpC or area
  subtraction. 
  One difference in behaviour that might explain this is that the
  $p_t$ threshold for cleansing's trimming step is
  $f_\text{cut} p_t^\text{full,no-CHS}$ (even in the CHS-like
  \texttt{input\_nc\_separate} mode that we use).
  In contrast, for the area and NpC-based results, it is
  $f_\text{cut} p_t^\text{full,CHS}$.
  In both cases the threshold, which is applied to subtracted subjets, is
  increased in the presence of pileup, but this increase is more
  substantial in the cleansing case.
  This could conceivably worsen the correspondence between trimming in
  the hard and full samples.
  For the area and NpC cases, we investigated the option of using
  $f_\text{cut} p_t^\text{area-sub,CHS}$ or
  $f_\text{cut} p_t^\text{NpC-sub,CHS}$ and found that this brings a
  small additional benefit. 
}
Does that mean that (protected) zeroing has no scope for improving the
trimmed-jet mass?
The answer is ``not necessarily'': one could for example imagine
first applying protected zeroing to subjets on some angular scale
$R_\text{zero}$ in order to eliminate low-$p_t$ contamination;
then reclustering the remaining constituents on a scale $R_\text{trim}
\gtrsim R_\text{zero}$, subtracting according to the area or NpC
methods, and finally applying the trimming momentum cut (while also
keeping in mind the considerations of footnote \ref{footnote:trimming-ref}).

We close this section with a summary of our findings.
Based on its description in Ref.~\cite{Krohn:2013lba} and our findings
about NpC v.\ area subtraction, cleansing with $f_\cut=0$ would be
expected to have a performance very similar to that of CHS+area
subtraction with $f_\cut=0$ trimming.
Ref.~\cite{Krohn:2013lba} however reported large improvements for the
correlation coefficients of the dijet mass and the single jet mass
using $R=1$ jets.
In the case of the dijet mass we do not see these improvements, though
they do appear to be there for the jet mass.
The differences in behaviour between cleansing and trimmed
CHS+area-subtraction call for an explanation, and appear to be
due to a step in the cleansing code that was undocumented in
Ref.~\cite{Krohn:2013lba} and that we dubbed ``zeroing'': if a subjet
contains no charged tracks from the leading vertex it is discarded.
Zeroing is an extreme form of a procedure described in
Ref.~\cite{ATL-Pileup-2}.
In can be used also with area or NpC subtraction, and we find that it
brings a benefit for the peak of the $\Delta p_t$ and $\Delta m$
distributions, but appears to introduce long tails in $\Delta p_t$.
A variant, ``protected zeroing'', can avoid the long tails by still
accepting subjets without leading-vertex tracks, if their $p_t$ is
above some threshold, which may be chosen dynamically based on the
properties of the pileup.
In our opinion, a phenomenologicaly realistic estimate of the benefit
of zeroing (protected or not) requires study not of $R=1$ plain jets,
but instead of $R=0.4$ jets (for the jet $p_t$) or larger-$R$ trimmed
jets with a non-zero $f_\cut$ (for the jet mass).
In a first investigation, there appear to be some phenomenological
benefits from protected zeroing for the $R=0.4$ jet $p_t$, whereas to
obtain benefits for large-$R$ trimmed jets would probably require
further adaptation of the procedure.
In any case, additional study is required for a full evaluation of
protected zeroing and related procedures.

%----------------------------------------------------------------------
\section{Combining NpC and area subtraction}
\label{sec:combination}

It is interesting to further probe the relation between NpC and the
area--median method, to establish whether there might be a benefit
from combining them:
the area--median method makes a mistake in predicting local energy flow
mainly because local energy flow fluctuates from region to region.
NpC makes a mistake because charged and neutral energy flow are not
$100\%$ locally correlated.
The key question is whether, for a given jet, NpC and the area--median
method generally make the same mistake, or if instead they are making
uncorrelated mistakes.
In the latter case it should be possible to combine the information
from the two methods to obtain an improvement in subtraction
performance.

\begin{figure}[t]
  \centering
  \includegraphics[width=0.48\textwidth]{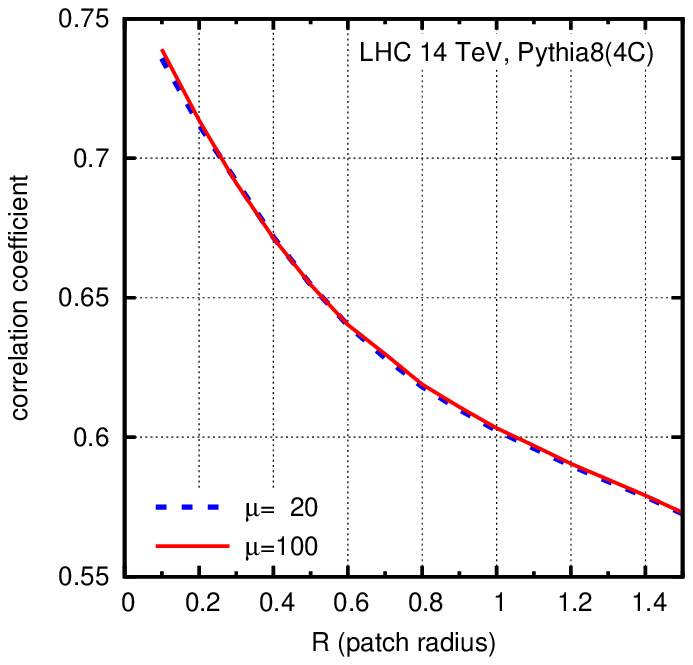}
  \hfill
  \includegraphics[width=0.48\textwidth]{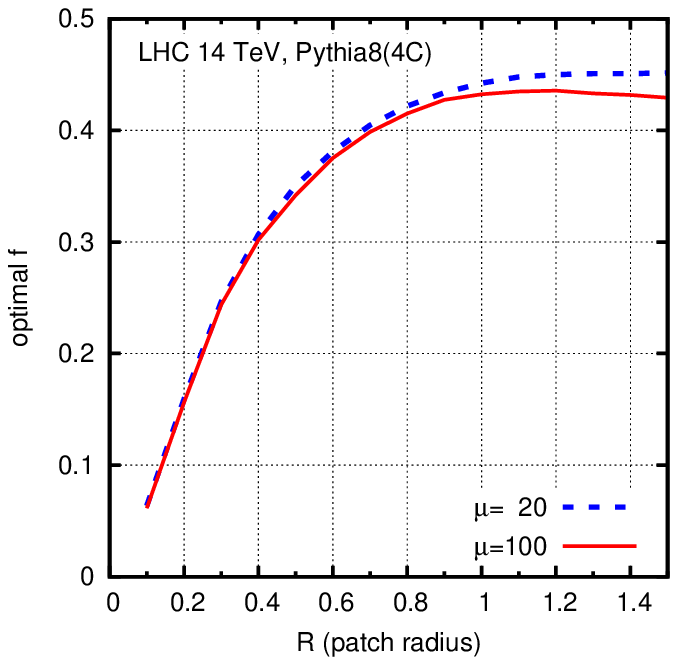}
  \caption{Left: correlation between $p_t^\ntr - \hat
    p_\mu^\text{ntr(NpC)}$ and $p_t^\ntr - \hat p_\mu^{\text{ntr($\rho
        A$)}}$, shown as a function of $R$. 
    Right: optimal weight $f$ for combining NpC and area pileup
    subtraction, Eq.~(\ref{eq:best f}), as a function of $R$.
  }
  \label{fig:correl-mistakes}
\end{figure}

Let $p_t^\ntr$ be the actual neutral pileup component flowing into a
jet, while
\begin{equation}
  \label{eq:pthat}
  \hat p_\mu^\text{ntr(NpC)} = \frac{1-\gamma_0}{\gamma_0 \epsilon}
  p_\mu^\text{jet,rescaled-chg-PU} \,, 
  \qquad 
  \hat p_\mu^{\text{ntr($\rho A$)}} = \rho A\,,
\end{equation}
are, respectively, the estimates for the neutral pileup based on the
local charged $p_t$ flow and on $\rho A$. We assume the use of CHS
events and, in particular, that $\rho$ is as determined from the CHS
event.
Concentrating on the transverse components, 
the extent to which the two estimates provide complementary
information can be quantified in terms of (one minus) the correlation
coefficient, $r$, between $p_t^\ntr - \hat p_\mu^\text{ntr(NpC)}$ and 
$p_t^\ntr - \hat p_\mu^{\text{ntr($\rho A$)}}$.
That correlation is shown as a function of $R$ in
Fig.~\ref{fig:correl-mistakes} (left), and it is quite high, in the range
$0.6$--$0.7$ for commonly used $R$ choices. 
It is largely independent
of the number of pileup vertices.

Let us now quantify the gain to be had from a linear combination of
the two prediction methods, i.e.\ using an estimate
\begin{equation}
  \label{eq:NpC+rhoA}
  \hat p_{\mu}^\text{ntr} = f \hat p_\mu^\text{ntr(NpC)} + (1-f) \hat p_\mu^{\text{ntr($\rho A$)}}\,,
\end{equation}
where $f$ is to be chosen to as to minimise the dispersion of
$p_t^\ntr - \hat p_t^\text{ntr}$.
Given dispersions $\sigma_\text{NpC}$ and $
\sigma_{\rho A}$
respectively for $p_t^\ntr - \hat p_\mu^\text{ntr(NpC)}$ and $p_t^\ntr - \hat
p_\mu^{\text{ntr($\rho A$)}}$, the optimal $f$ is
\begin{equation}
  \label{eq:best f}
  f = \frac{\sigma_{\rho A}^2 - r \, \sigma_{\NpC} \, \sigma_{\rho
      A}}{\sigma_{\NpC}^2 +  \sigma_{\rho A}^2 - 2 r \,\sigma_{\NpC} \, \sigma_{\rho A}}\,,
\end{equation}
which is plotted as a function of $R$ in
Fig.~\ref{fig:correl-mistakes} (right), and the resulting squared dispersion
for $p_t^\ntr - \hat p_t^\text{ntr}$ is
\begin{equation}
  \label{eq:combined-dispersion}
  \sigma^2
  =
  \frac{(1-r^2) \, \sigma_{\NpC}^2\, \sigma_{\rho
      A}^2}{\sigma_{\NpC}^2 + \sigma_{\ntr}^2 - 2 r\, \sigma_{\NpC}\, \sigma_{\rho A}}\,.
\end{equation}
Reading $r = 0.67$ from Fig.~\ref{fig:correl-mistakes} (left) for $R=0.4$, and
$ \sigma_{\NpC} \simeq 1.14\, \sigma_{\rho A}$ from
Fig.~\ref{fig:correl-central} (right), one finds $\sigma \simeq 0.96 \,
\sigma_{\rho A}$.
Because of the substantial correlation between the two methods, ones
expects only a modest gain from their linear combination.

\begin{figure}
  \centering
  \includegraphics[width=0.48\textwidth]{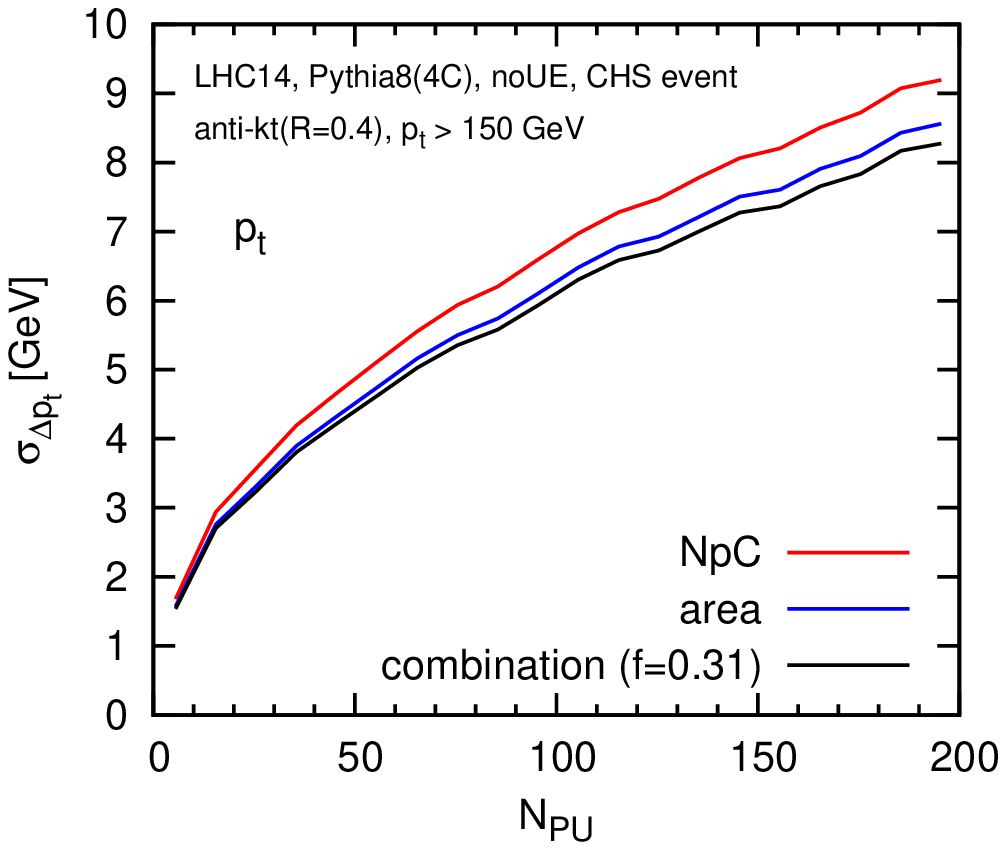}
  \hfill
  \includegraphics[width=0.48\textwidth]{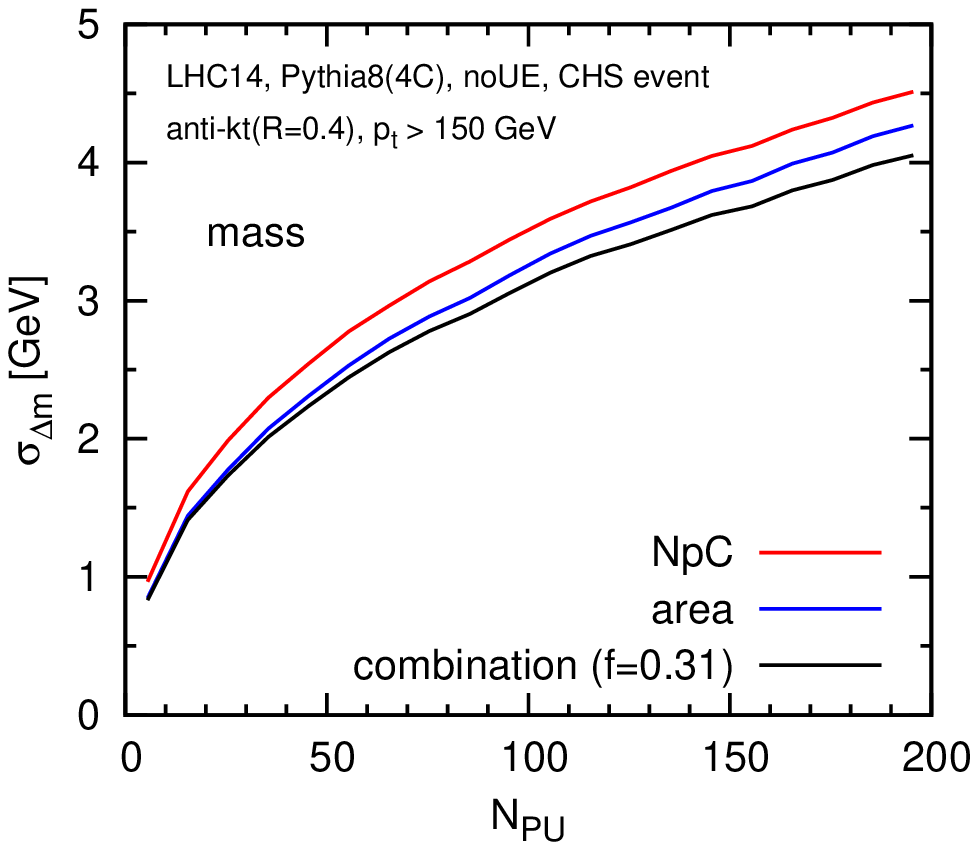}
  \caption{Comparison of the performance of NpC, area--median and
    combined subtraction, as a function of the number of pileup
    vertices. The left-hand plot is for the jet $p_t$ and the
    right-hand one for the jet mass.}
  \label{fig:chg+rhoA-improvement}
\end{figure}

In Fig.~\ref{fig:chg+rhoA-improvement} we compare the performance of
pileup subtraction from the combination of the NpC and the
area--median methods, using the optimal value $f=0.31$ that can be read from
Fig.~\ref{fig:correl-mistakes} (right) for $R=0.4$, both for the jet $p_t$ and the
jet mass.
The expected small gain is indeed observed for the jet $p_t$, and it
is slightly larger for the jet mass.\footnote{We also examined
  results with other choices for $f$: we found that the true optimum
  value of $f$ in the Monte Carlo studies is slightly different from
  that predicted by Eq.~(\ref{eq:best f}). However the dependence on
  $f$ around its minimum is very weak, rendering the details of its
  exact choice somewhat immaterial. }
Given the modest size of the gain, one may wonder how
phenomenologically relevant it is likely to be. 
Nevertheless, one might still consider investigating whether the gain
carries over also to a realistic experimental environment with full
detector effects.

%======================================================================
\section{Conclusions}
\label{sec:concl}

One natural approach to pileup removal is to use the charged pileup
particles in a given jet to estimate the amount of neutral pileup that
needs to be removed from that same jet.
In this article, with the help of particle-level simulations, we have
studied such a method (NpC) and found that it has a performance that
is similar to, though slightly worse than the existing, widely used
area--median method.
This can be related to the observation that the correlations between
local charged and neutral energy flow are no larger than those between
global and local energy flow.
Tentatively, we believe that this is in part because the non-perturbative
effects that characterise typical inelastic proton-proton collisions
act to destroy local charged-neutral correlation.

The absence of benefit that we found from the NpC method led us to
question the substantial performance gains quoted for the method of
cleansing in Ref.~\cite{Krohn:2013lba},
one of whose key differences with respect to earlier work is the
replacement of the area--median method with NpC.
For the dijet mass, we are unable to reproduce the large improvement
observed in Ref.~\cite{Krohn:2013lba}, in the correlation coefficient
performance measure, for cleansing relative to area subtraction.
We do however see an improvement for the jet mass.
We trace a key difference in the behaviour of cleansing and area
subtraction to the use in the cleansing code of a step that was not
documented in Ref.~\cite{Krohn:2013lba} and that discards subjets that
contain no tracks from the leading vertex. 
This ``zeroing'' step, similar to the charged-track based trimming
introduced by ATLAS~\cite{ATL-Pileup-2}, can indeed be of benefit. 
It has a drawback of introducing tails in some distributions due to
subjets with a substantial neutral $p_t$ from the leading vertex, but
no charged tracks.
As a result, different quality measures lead to different conclusions
as to the benefits of zeroing.
The tails can be alleviated by a variant of zeroing that we introduce here,
``protected zeroing'', whereby subjets without LV charged tracks are
rejected only if their $p_t$ is below some (possibly pileup-dependent)
threshold.
Protected zeroing does in some cases appear to have phenomenological
benefits, which are observed across all quality
measures. 

Given two different methods for pileup removal, NpC and area--median
subtraction, it is natural to ask how independent they are and what
benefit might be had by combining them.
This was the question investigated in section~\ref{sec:combination},
where we provided a formula for an optimal linear combination of the
two methods, as a function of their degree of correlation.
Ultimately we found that NpC and area--median subtraction are quite
highly correlated, which limits the gains from their combination to
about a $5\%$ percent reduction in dispersion.
While modest, this might still be sufficient to warrant experimental
investigation, as are other methods, currently being developed, that
exploit constituent-level
subtraction~\cite{YueShi,Berta:2014eza,Cacciari:2014gra,Bertolini:2014bba}. 
A study of the integration of those methods with protected zeroing
would also be of interest.

Code for our implementation of area subtraction with positive-definite
mass is available as part of FastJet versions 3.1.0 and higher.
Public code and samples for carrying out a subset of the comparisons
with cleansing described in section~\ref{sec:appraisal}, including
also the NpC subtraction tools, are available from
Ref.~\cite{public-code}.

\section*{Acknowledgements}
Our understanding of pileup effects in the LHC experiments has
benefited extensively from discussions with Peter Loch, David Miller,
Filip Moortgat, Sal Rappoccio, Ariel Schwartzman and numerous others.
We are grateful to David Krohn, Matthew Low, Matthew Schwartz and
Liantao Wang for exchanges about their results.
This work was supported by ERC advanced grant Higgs@LHC, by the French
Agence Nationale de la Recherche, under grant ANR-10-CEXC-009-01, 
by the EU ITN grant LHCPhenoNet, PITN-GA-2010-264564 and by the ILP LABEX
(ANR-10-LABX-63) supported by French state funds
managed by the ANR within the Investissements d'Avenir programme
under reference ANR-11-IDEX-0004-02.
GPS wishes to thank Princeton University for hospitality while this
work was being carried out.
GS wishes to thank CERN for hospitality while this work was being
finalised.

\appendix
%----------------------------------------------------------------------
\section{Details of our study}
\label{sec:details}

Let us first fully specify what we have done in our study and then
comment on (possible) differences relative to KLSW.

Our hard event sample consists of dijet events from $pp$ collisions at
$\sqrt{s}=14\TeV$, simulated with Pythia~8.176~\cite{Pythia8}, tune
4C, with a minimum $p_t$ in the $2\to2$ scattering of $135\GeV$ and
with the underlying event turned off, except for the plots presented
in Figs.~\ref{fig:f00} and \ref{fig:shifts-dispersions-R1},
where we 
use $Z'$ events with $m_{Z'}=500\GeV$.
Jets are reconstructed with the anti-$k_t$
algorithm~\cite{Cacciari:2008gp} after making all particles massless
(preserving their rapidity) and keeping only particles with
$|y|<4$. We have $R=0.4$, except for the some of the results presented in
Section~\ref{sec:appraisal} and Appendix~\ref{sec:correlation-coefs},
where we use $R=1$ as in 
Ref.~\cite{Krohn:2013lba}.

Given a hard event, we select all the jets with $p_t > 150\GeV$ and
absolute rapidity $|y| < 2.5$. We then add pileup and cluster
the resulting full event,  i.e.\ including both the hard event and
the pileup particles, without imposing any $p_t$ or rapidity cut on
the resulting jets. For each jet selected in the hard event as
described above, we find the jet in the full event that overlaps the
most with it. 
Here, the overlap is defined as the scalar $p_t$ sum of
all the common jet constituents, as described in
footnote~\ref{footnote:matching} on p.~\pageref{footnote:matching}.
Given a pair of jets, one in the hard
event and the matching one in the full event, we can apply
subtraction/grooming/cleansing to the latter and study the quality of
the jet $p_t$ or jet mass reconstruction.
For studies involving the dijet mass (cf. Fig.~\ref{fig:f00})
we require that at least two jets pass the jet selection in the
hard event and use those two hardest jets, and the corresponding
matched ones in the full event, to reconstruct the dijet
mass.\footnote{In the case of the $Z'$ events used for
  Fig.~\ref{fig:f00}, this does not exactly reflect how we
  would have chosen to perform a dijet (resonance) study ourselves.
  One crucial aspect is that searches for dijet resonances always
  impose a rapidity cut between the two leading jets, such as $|\Delta
  y| < 1.2$~\cite{ATLASDijet,CMSDijet}. This ensures that high
  dijet-mass events are not dominated by low $p_t$ forward-backward
  jet pairs, which are usually enhanced in QCD v.\ resonance
  production.
  Those forward-backward pairs can affect conclusions about pileup,
  because for a given dijet mass the jet $p_t$'s in a forward-backward
  pair are lower than in a central-central pair, and so relatively
  more sensitive to pileup.
  Also the experiments do not use $R=1$ for their dijet studies: ATLAS
  uses $R=0.6$~\cite{ATLASDijet}, while CMS uses $R=0.5$ with a form of
  radiation recovery based on the inclusion of any additional jets
  with $p_t > 30 \GeV$ and within $\Delta R = 1.1$ of either of the
  two leading jets (``wide jets'')~\cite{CMSDijet}.
  This too can affect conclusions about pileup.
}
This approach avoids having to additionally consider the impact of
pileup on the efficiency for jet selection, which cannot
straightforwardly be folded into our quality measures.\footnote{One
  alternative would have been to impose the cuts on the jets in the
  full event (with pileup and subtraction/grooming/cleansing) and
  consider as the ``hard jet'', the subset of the particles in the
  full-event jet that come from the leading vertex ({\it i.e.} the
  hard event).
  We understand that this is close to the choice made in
  Ref.~\cite{Krohn:2013lba}. 
  This can give overly optimistic results, because it neglects
  backreaction. However in our studies it did not appear to
  significantly modify the conclusions on relative performances of
  different methods.
}

Most of the studies shown in this paper use idealised particle-level
CHS events. 
In these events, we scale all charged pileup hadrons by
a factor $\epsilon=10^{-60}$ before clustering, to ensure that they do
not induce any backreaction~\cite{areas}. The jet selection and
matching procedures are independent of the use of CHS or full events.
When we plot results as a function of the quantity $N_\text{PU}$, this
corresponds to the actual (fixed) number of zero-bias events
superimposed onto 
the hard collision. For results shown as a function of $\mu$, the
average number of zero-bias events, the actual number of zero-bias
events has a Poisson distribution.
Clustering and area determination are performed with a development
version of FastJet 3.1 (which for the features used here behaves
identically to the 3.0.x series) and with FastJet 3.1.0 and 3.1.1 for
the results in section~\ref{sec:appraisal}.

Details of how the area-median subtraction is performed could
conceivably matter.
Jet areas are obtained using active area with explicit ghosts placed
up to $|y|=4$ and with a default ghost area of 0.01. We use FastJet's
\texttt{GridMedianBackgroundEstimator} with a grid spacing of 0.55 to
estimate the event background density $\rho$. The $\rho$ estimation is
performed using the particles (up to $|y| = 4$) from the full
or the CHS event as appropriate.
When subtracting pileup from jets, we account for the rapidity
dependence of $\rho$, based on the rapidity dependence in a pure
pileup sample (as discussed in Refs.~\cite{FastJet,AlcarazMaestre:2012vp}).
We carry out 4-vector subtraction $p_\mu^{\text{sub}} = p_\mu - A_\mu
\rho(y_\text{jet})$.

A few obviously unphysical situations need special care. For jets
obtained from the full event, if $A_t \rho > p_{t}^{\text{jet}}$, we
set $p_\mu^\text{sub}$ to a vector with zero transverse momentum, zero
mass, and the rapidity and azimuth of the original unsubtracted jet;
and if $(p^\text{sub})^2$ is negative, an unphysical situation since
it would lead to an imaginary mass, we replace $p_\mu^\text{sub}$ with
a vector with the same transverse components, zero mass, and the
rapidity of the original unsubtracted jet.\footnote{In versions of
  FastJet prior to 3.1.0 this had to be done manually: the treatment
  of unphysical 4-vectors was left to the user, since the optimal
  treatment may depend on the context.
  As of version 3.1.0, FastJet provides the option of ``safe''
  subtraction, whereby subjets with negative mass squared after
  subtraction are automatically assigned
  a zero mass (in CHS events, the safe procedure follows the
  description below in the text).
}
This is essentially equivalent to replacing negative squared masses
with zero.

The case of CHS events is a bit more delicate. Let
$p_\mu^{\text{chg}}$ denote the 4-momentum of the charged component of
the jet.\footnote{For our ``emulated'' CHS events, the charged
  contribution from pileup is 
  negligible since it has been scaled down, and only the contribution
  from the hard interaction counts.} Then, if
$p_{t}^{\text{sub}}<p_{t}^{\text{chg}}$, we set
$p_\mu^\text{sub}=p^{\text{chg}}_\mu$, and when
$(p^{\text{sub}})^2<(p^{\text{chg}})^2$, we replace $p_\mu^\text{sub}$
with a vector with the same transverse components, and the mass and
rapidity of $p^{\text{chg}}_\mu$. For jets with no charged component,
whenever the resulting 4-vector has an ill-defined rapidity or
azimuthal angle, we use those of the original jet.
Corresponding tests that the subtracted transverse momentum and mass
are non-negative are also applied in our NpC subtraction.
These safety requirements have little impact on the single-jet $p_t$,
limited impact on the dijet mass, and for the single-jet mass improve
the dispersion of the subtraction relative to the choice (widespread
in computer codes) of taking $m\equiv -\sqrt{|m^2|}$ when $m^2 <
0$.\footnote{As this work was being finalised, an alternative approach
  to using area--median information to reconstruct the 4-vector (and
  shapes) of a jet was proposed in Ref.~\cite{Berta:2014eza}, based on
  a constituent-level assignment of the subtraction. It appears to
  bring further improvements in performance for observables like the
  jet mass. }

One difference between our study and KLSW's is that we carry out a
particle-level study, whereas they project their event onto a toy
detector with a $0.1\times0.1$ tower granularity, removing charged
particles with $p_t < 0.5\GeV$ and placing a $1\GeV$ threshold on
towers.
In our original (v1) studies with $f_\cut=0.05$ we tried including a
simple detector simulation along these lines and did not find any
significant modification to the pattern of our results, though
CHS+area subtraction is marginally closer to the cleansing curves in
this case.\footnote{We choose to show particle-level results here
  because of the difficulty of correctly simulating a full detector,
  especially given the non-linearities of responses to pileup and the
  subtleties of particle flow reconstruction in the presence of real
  detector fluctuations.}

Cleansing has two options: one can give it jets clustered from the
full event, and then it uses an analogue of Eq.~(\ref{eq:npc-full}):
this effectively subtracts the exact charged part and the NpC estimate
of the neutrals.
Or one can give it jets clustered from CHS events, and it then applies
the analogue of Eq.~(\ref{eq:npc-chs}), which assumes that there is no
charged pileup left in the jet and uses just the knowledge of the
actual charged pileup to estimate (and subtract) the neutral pileup.
These two approaches differ by contributions related to back-reaction.
Our understanding is that KLSW took the former approach, while we used
the latter. 
Specifically, our charged-pileup hadrons, which are scaled down in the CHS event,
are scaled back up to their original $p_t$ before passing them to the
cleansing code, in its \texttt{input\_nc\_separate} mode.
If we use cleansing with full events, we find that its performance
worsens, as is to be expected given the additional backreaction
induced when clustering the full event. Were it not for backreaction,
cleansing applied to full or CHS events should essentially be
identical. 

Regarding the NpC and cleansing parameters, our value of $\gamma_0 =
0.612$ differs slightly from that of KLSW's $\gamma_0=0.55$, and
corresponds to the actual fraction of charged pileup in our simulated
events.
In our tests with a detector simulation like that of KLSW, we
adjusted $\gamma_0$ to its appropriate (slightly lower) value.

Finally, for trimming we use $R_{\rm trim}=0.3$ and the
reference $p_t$ is taken unsubtracted, while the subjets are
subtracted before performing the trimming cut, which removes subjets
with $p_t$ below a fraction $f_\text{cut}$ times
the reference $p_t$.
Compared to using the subtracted $p_t$ as the reference for trimming,
this effectively places a somewhat harder cut as pileup is
increased.\footnote{And is the default behaviour in FastJet if one
  passes an unsubtracted jet to a trimmer with subtraction, e.g.\ a
  \texttt{Filter(Rtrim,SelectorPtFractionMin(fcut),$\rho$)}. 
  One may of course choose to pass a subtracted jet to the trimmer, in
  which case the reference $p_t$ will be the subtracted one.
}
For comparisons with cleansing we generally use $f_\cut = 0$ unless
explicitly indicated otherwise.

%----------------------------------------------------------------------
\section{The correlation coefficient as a quality measure}
\label{sec:correlation-coefs}

In this appendix, we discuss some characteristics of correlation
coefficients that affect their appropriateness as generic quality
measures for pileup studies.

Suppose we have an observable $v$.
Define 
\begin{equation}
  \Delta v = v^\text{sub} - v^\text{hard}\,,
\end{equation}
to be the difference, in a given event, between the pileup subtracted
observable and the original ``hard'' value without pileup.
Two widely used
quality measures for the performance of pileup subtraction are the
average offset of $v$, $\langle \Delta v\rangle$ and the standard
deviation of $\Delta v$, which we write as $\sigma_{\Delta v}$.
One might think there is a drawback in keeping track of two measures,
in part because it is not clear which of the two is more important.
It is our view that the two measures provide complementary
information: if one aims to reduce systematic errors in a precision
measurement then a near-zero average offset may be the most important
requirement, so as not to be plagued by issues related to the
systematic error on the offset.
In a search for a resonance peak, then one aims for the narrowest
peak, and so the smallest possible standard deviation.\footnote{This
  statement assumes the absence of tails in the $\Delta v$
  distribution. 
  For some methods the long tails can affect the relevance of the
  standard-deviation quality measure.
}

The quality measure advocated in \cite{Krohn:2013lba} is instead the
correlation coefficient between $v^\text{sub}$
and $v^\text{hard}$.
This has the apparent simplifying advantage of providing just a single
quality measure. 
However, it comes at the expense of masking potentially important
information: for example, a method with a large offset and one with no
offset will give identical correlation coefficients, because the
correlation coefficient is simply insensitive to (constant)
offsets.

\begin{figure}[t]
  \centering
  \includegraphics[width=0.48\textwidth]{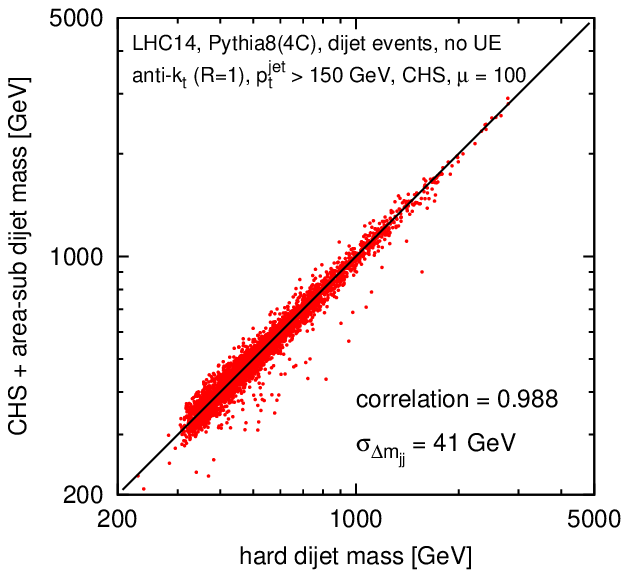}\hfill
  \includegraphics[width=0.48\textwidth]{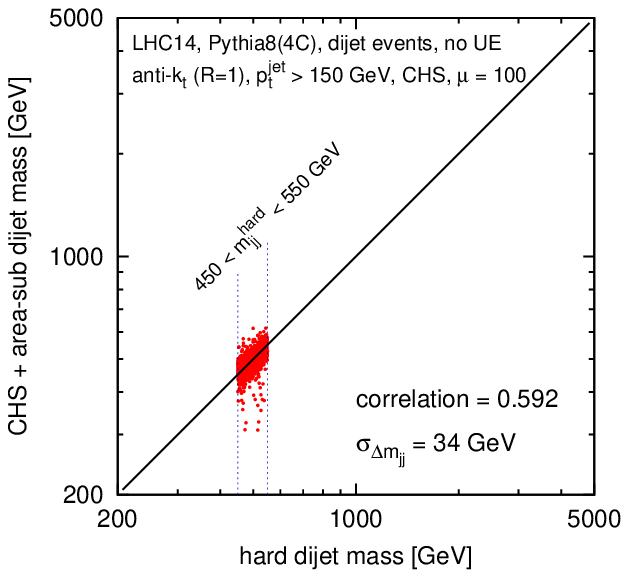}
  \caption{Left: scatter plot of the dijet mass after addition of an
    average of $100$ pileup 
    events and area--median subtraction (in CHS events) versus the dijet mass in the original
    hard event. 
    The hard dijet sample and the analysis are as described in
    appendix~\ref{sec:details}, with a jet radius of $R=1$.
    The right-hand plot is identical except for the following
    additional condition on the hard event: $450 < m_{jj} < 550
    \GeV$. Note the lower correlation coefficient, even though the
    lower $\sigma_{\Delta m_{jj}}$ suggests better typical subtraction in
    this specific mass bin. }
  \label{fig:correls-are-bad}
\end{figure}

The correlation coefficient has a second, more fundamental flaw, as
illustrated in Fig.~\ref{fig:correls-are-bad}. On the left, one has a
scatter plot of the dijet mass in PU-subtracted events versus the
dijet mass in the corresponding hard events, as obtained in an
inclusive jet sample.
There is a broad spread of dijet masses, much wider than the standard
deviation of $\Delta m_{jj}$, and so the correlation coefficient
comes out very high, $c= 0.988$.
Now suppose we are interested in reconstructing resonances with a mass
near $500\GeV$, and so consider only hard events in which $450 <
m_{jj} < 550\GeV$ (right-hand plot).
Now the correlation coefficient is $0.59$, i.e.\ much
worse.
This does not reflect a much worse subtraction: actually,
$\sigma_{\Delta m_{jj}}$ is better (lower) in the sample with a limited
$m_{jj}$ window, $\sigma_{\Delta m_{jj}}=34\GeV$, than in the full sample,
$\sigma_{\Delta m_{jj}}=41\GeV$.
The reason for the puzzling decrease in the correlation coefficient is
that the dispersion of $m_{jj}$ is much smaller than before, and so
the dispersion of $\Delta m_{jj}$ is now comparable to that of
$m_{jj}$: it is this, and not an actual degradation of performance,
that leads to a small correlation.

This can be understood quantitatively in a simple model with two
variables: let $v^\text{hard}$ have a standard deviation of
$\sigma_{v,\text{hard}}$, and
for a given $v^\text{hard}$ let $v^\text{sub}$ be distributed with a mean value
equal to $v^\text{hard}$ (plus an optional constant offset) and a
standard deviation of $\sigma_{\Delta v}$ (independent of $v^\text{hard}$).
Then the correlation coefficient of $v^\text{hard}$ and
$v^\text{sub}$ is
\begin{equation}
  \label{eq:correl-model}
  c = \frac{\sigma_{v,\text{hard}}}{\sqrt{(\sigma_{v,\text{hard}})^2 +
      \sigma_{\Delta v}^2}}\,,
\end{equation}
i.e.\ it tends to zero for $\sigma_{v,\text{hard}} \ll \sigma_{\Delta
  v}$ and to $1$ for large $\sigma_{v,\text{hard}} \gg \sigma_{\Delta
  v}$, in accord with the qualitative behaviour seen in
Fig.~\ref{fig:correls-are-bad}.
The discussion becomes more involved if $\langle v^\text{sub}\rangle$
has a more complicated dependence on $v^\text{hard}$ or if
$\sigma_{\Delta v}$ itself depends on $v^\text{hard}$, for example as
is actually the case for the dijet mass with the analysis of
Appendix~\ref{sec:details}.

The main conclusion from this appendix is that correlation
coefficients mix together information about the quality of pileup
mitigation and information about the hard event sample being studied.
It is then highly non-trivial to extract just the information about
the pileup subtraction.
This can lead to considerable confusion, for example, when evaluating
the robustness of a method against the choice of hard sample.
Overall therefore, it is our recommendation that one consider direct
measures of the dispersion introduced by the pileup and subtraction
and not correlation coefficients.
In cases with severely non-Gaussian tails in the $\Delta v$
distributions it can additionally be useful to consider quality
measures more directly related to the peak structure of the $\Delta v$
distribution.


\begin{thebibliography}{99}

\bibitem{ATLAS-PU-Performance}
  The ATLAS collaboration,
  ``Pile-up subtraction and suppression for jets in ATLAS,''
  \href{http://inspirehep.net/record/1260963/files/ATLAS-CONF-2013-083.pdf}{ATLAS-CONF-2013-083}.
  %%CITATION = ATLAS-CONF-2013-083;%%
  %13 citations counted in INSPIRE as of 25 Mar 2014

\bibitem{CMS-PU-Performance}
  CMS Collaboration,
  ``Jet Energy Scale performance in 2011,''
  \href{http://inspirehep.net/record/1230033/files/DP2012\_006.pdf}{CMS-DP-2012-006}.
  %%CITATION = CMS-DP-2012-006;%%
  %4 citations counted in INSPIRE as of 25 Mar 2014

\bibitem{areasub}
  M.~Cacciari and G.~P.~Salam,
  %``Pileup subtraction using jet areas,''
  Phys.\ Lett.\ B {\bf 659} (2008) 119
  [arXiv:0707.1378 [hep-ph]].
  %%CITATION = ARXIV:0707.1378;%%
  %311 citations counted in INSPIRE as of 11 Dec 2013

\bibitem{areas}
  M.~Cacciari, G.~P.~Salam and G.~Soyez,
  %``The Catchment Area of Jets,''
  JHEP {\bf 0804} (2008) 005
  [arXiv:0802.1188 [hep-ph]].
  %%CITATION = ARXIV:0802.1188;%%
  %245 citations counted in INSPIRE as of 11 Dec 2013

\bibitem{pflow}
  CMS Collaboration,
  ``Particle-Flow Event Reconstruction in CMS and Performance for Jets, Taus, and MET,''
  CMS-PAS-PFT-09-001.
  %%CITATION = CMS-PAS-PFT-09-001;%%
  %341 citations counted in INSPIRE as of 11 Dec 2013

\bibitem{Aad:2012tfa}
  G.~Aad {\it et al.}  [ATLAS Collaboration],
  %``Observation of a new particle in the search for the Standard Model Higgs boson with the ATLAS detector at the LHC,''
  Phys.\ Lett.\ B {\bf 716} (2012) 1
  [arXiv:1207.7214 [hep-ex]].
  %%CITATION = ARXIV:1207.7214;%%
  %2368 citations counted in INSPIRE as of 29 Mar 2014

\bibitem{Colas:2005jn}
  J.~Colas {\it et al.}  [ATLAS Liquid Argon Calorimeter Collaboration],
  %``Position resolution and particle identification with the ATLAS EM calorimeter,''
  Nucl.\ Instrum.\ Meth.\ A {\bf 550} (2005) 96
  [physics/0505127].
  %%CITATION = PHYSICS/0505127;%%
  %32 citations counted in INSPIRE as of 16 Apr 2014


\bibitem{ourtalk4cms}
  M.~Cacciari, G.~P.~Salam and G.~Soyez, unpublished, presented at
  CMS Week, CERN, Geneva, Switzerland, March 2011, available publicly
  since then at
  \url{http://www.lpthe.jussieu.fr/~salam/talks/repo/2011-CMS-week.pdf}. 

\bibitem{Krohn:2013lba}
  D.~Krohn, M.~D.~Schwartz, M.~Low and L.~T.~Wang,
  %``Jet Cleansing: Pileup Removal at High Luminosity,''
  Phys.\ Rev.\ D {\bf 90} (2014) 6,  065020
  [arXiv:1309.4777 [hep-ph]].
  %%CITATION = ARXIV:1309.4777;%%

\bibitem{CMS:2013wea}
  CMS Collaboration,
  ``Pileup Jet Identification,''
  \href{http://cds.cern.ch/record/1581583?ln=en}{CMS-PAS-JME-13-005}.
  %%CITATION = CMS-PAS-JME-13-005;%%
  %9 citations counted in INSPIRE as of 29 Mar 2014

\bibitem{ATL-Pileup-2}
  ATLAS Collaboration,
  ``Tagging and suppression of pileup jets,''
  % includes cleansing
  \href{https://cds.cern.ch/record/1643929?ln=en}{ATL-PHYS-PUB-2014-001}.

\bibitem{quality}
  M.~Cacciari, J.~Rojo, G.~P.~Salam and G.~Soyez,
  %``Quantifying the performance of jet definitions for kinematic reconstruction at the LHC,''
  JHEP {\bf 0812} (2008) 032
  [arXiv:0810.1304 [hep-ph]].
  %%CITATION = ARXIV:0810.1304;%%
  %34 citations counted in INSPIRE as of 11 Dec 2013

\bibitem{boost2012}
  A.~Altheimer, A.~Arce, L.~Asquith, J.~Backus Mayes, E.~Bergeaas Kuutmann, J.~Berger, D.~Bjergaard and L.~Bryngemark {\it et al.},
  %``Boosted objects and jet substructure at the LHC,''
  arXiv:1311.2708 [hep-ex].
  %%CITATION = ARXIV:1311.2708;%%
  %2 citations counted in INSPIRE as of 11 Dec 2013

\bibitem{filter}
  J.~M.~Butterworth, A.~R.~Davison, M.~Rubin and G.~P.~Salam,
  %``Jet substructure as a new Higgs search channel at the LHC,''
  Phys.\ Rev.\ Lett.\  {\bf 100} (2008) 242001
  [arXiv:0802.2470 [hep-ph]].
  %%CITATION = ARXIV:0802.2470;%%
  %356 citations counted in INSPIRE as of 11 Dec 2013

\bibitem{trim}
  D.~Krohn, J.~Thaler and L.~-T.~Wang,
  %``Jet Trimming,''
  JHEP {\bf 1002} (2010) 084
  [arXiv:0912.1342 [hep-ph]].
  %%CITATION = ARXIV:0912.1342;%%
  %123 citations counted in INSPIRE as of 11 Dec 2013

\bibitem{Cacciari:2010te}
  M.~Cacciari, J.~Rojo, G.~P.~Salam and G.~Soyez,
  %``Jet Reconstruction in Heavy Ion Collisions,''
  Eur.\ Phys.\ J.\ C {\bf 71} (2011) 1539
  [arXiv:1010.1759 [hep-ph]].
  %%CITATION = ARXIV:1010.1759;%%
  %36 citations counted in INSPIRE as of 30 Mar 2014

\bibitem{Soyez:2012hv}
  G.~Soyez, G.~P.~Salam, J.~Kim, S.~Dutta and M.~Cacciari,
  %``Pileup subtraction for jet shapes,''
  Phys.\ Rev.\ Lett.\  {\bf 110} (2013) 16,  162001
  [arXiv:1211.2811 [hep-ph]].
  %%CITATION = ARXIV:1211.2811;%%
  %20 citations counted in INSPIRE as of 07 Feb 2014

\bibitem{Cacciari:2012mu}
  M.~Cacciari, P.~Quiroga-Arias, G.~P.~Salam and G.~Soyez,
  %``Jet Fragmentation Function Moments in Heavy Ion Collisions,''
  Eur.\ Phys.\ J.\ C {\bf 73} (2013) 2319
  [arXiv:1209.6086 [hep-ph]].
  %%CITATION = ARXIV:1209.6086;%%
  %3 citations counted in INSPIRE as of 04 Apr 2014
 
\bibitem{Cacciari:2008gp}
  M.~Cacciari, G.~P.~Salam and G.~Soyez,
  %``The Anti-k(t) jet clustering algorithm,''
  JHEP {\bf 0804} (2008) 063
  [arXiv:0802.1189 [hep-ph]].
  %%CITATION = ARXIV:0802.1189;%%
  %1855 citations counted in INSPIRE as of 19 Dec 2013

\bibitem{FastJet}
  M.~Cacciari, G.~P.~Salam and G.~Soyez,
  %``FastJet User Manual,''
  Eur.\ Phys.\ J.\ C {\bf 72} (2012) 1896
  [arXiv:1111.6097 [hep-ph]].
  %%CITATION = ARXIV:1111.6097;%%
  %383 citations counted in INSPIRE as of 11 Dec 2013

\bibitem{Pythia8}
  T.~Sjostrand, S.~Mrenna and P.~Z.~Skands,
  %``A Brief Introduction to PYTHIA 8.1,''
  Comput.\ Phys.\ Commun.\  {\bf 178} (2008) 852
  [arXiv:0710.3820 [hep-ph]].
  %%CITATION = ARXIV:0710.3820;%%
  %944 citations counted in INSPIRE as of 11 Dec 2013

\bibitem{seymour}
  M.~H.~Seymour,
  %``Searches for new particles using cone and cluster jet algorithms: A Comparative study,''
  Z.\ Phys.\ C {\bf 62} (1994) 127.
  %%CITATION = ZEPYA,C62,127;%%
  %109 citations counted in INSPIRE as of 11 Dec 2013


\bibitem{schwartz-talk-PU-workshop} 
  M.~D.~Schwartz, 
  %
  Talk at Workshop on Mitigation of Pileup Effects at the LHC, 
  CERN, May 2014,\\
  \url{https://indico.cern.ch/event/306155/session/2/contribution/1/material/slides/0.pdf}

\bibitem{fjcontrib}
  \url{http://fastjet.hepforge.org/contrib/}

\bibitem{CMS:2014ata}
  CMS Collaboration,
  ``Pileup Removal Algorithms,''
  CMS-PAS-JME-14-001.
  %%CITATION = CMS-PAS-JME-14-001;%%
  %2 citations counted in INSPIRE as of 16 Dec 2014


\bibitem{Kodolova}
  %
  O.L. Kodolova {\it et al}, 
  ``Study of $\gamma$+Jet Channel in Heavy ion Collisions with CMS,''
  CMS-NOTE-1998-063;\\
  %%CITATION = CMS-NOTE-1998-063;%%
  %
  V.~Gavrilov, A.~Oulianov, O.~Kodolova and I.~Vardanian, 
  ``Jet Reconstruction with Pileup Subtraction,''
  CMS-RN-2003-004;\\
  %%CITATION = CMS-RN-2003-004;%%
  %
  O.~Kodolova, I.~Vardanian, A.~Nikitenko and A.~Oulianov,
  %``The performance of the jet identification and reconstruction in heavy ions collisions with CMS detector,''
  Eur.\ Phys.\ J.\ C {\bf 50} (2007) 117.
  %%CITATION = EPHJA,C50,117;%%
  %39 citations counted in INSPIRE as of 27 May 2014 
%

\bibitem{Cacciari:2014gra}
  M.~Cacciari, G.~P.~Salam and G.~Soyez,
  %``SoftKiller, a particle-level pileup removal method,''
  Eur.\ Phys.\ J.\ C {\bf 75} (2015) 2,  59
  [arXiv:1407.0408 [hep-ph]].
  %%CITATION = ARXIV:1407.0408;%%
  %5 citations counted in INSPIRE as of 14 Feb 2015

\bibitem{Bertolini:2014bba}
  D.~Bertolini, P.~Harris, M.~Low and N.~Tran,
  %``Pileup Per Particle Identification,''
  JHEP {\bf 1410} (2014) 59
  [arXiv:1407.6013 [hep-ph]].
  %%CITATION = ARXIV:1407.6013;%%
  %5 citations counted in INSPIRE as of 05 Jan 2015

\bibitem{Dasgupta:2013ihk}
  M.~Dasgupta, A.~Fregoso, S.~Marzani and G.~P.~Salam,
  %``Towards an understanding of jet substructure,''
  JHEP {\bf 1309} (2013) 029
  [arXiv:1307.0007 [hep-ph]].
  %%CITATION = ARXIV:1307.0007;%%
  %13 citations counted in INSPIRE as of 07 Mar 2014

\bibitem{Dasgupta:2013via}
  M.~Dasgupta, A.~Fregoso, S.~Marzani and A.~Powling,
  %``Jet substructure with analytical methods,''
  Eur.\ Phys.\ J.\ C {\bf 73} (2013) 11,  2623
  [arXiv:1307.0013 [hep-ph]].
  %%CITATION = ARXIV:1307.0013;%%
  %8 citations counted in INSPIRE as of 04 Apr 2014


\bibitem{YueShi}
  CMS Collaboration,
  ``Underlying Event Subtraction for Particle Flow,''
  {\small \url{https://twiki.cern.ch/twiki/bin/view/CMSPublic/PhysicsResultsDPSUESubtractionPF}}

\bibitem{Berta:2014eza}
  P.~Berta, M.~Spousta, D.~W.~Miller and R.~Leitner,
  %``Particle-level pileup subtraction for jets and jet shapes,''
  JHEP {\bf 1406} (2014) 092
  [arXiv:1403.3108 [hep-ex]].
  %%CITATION = ARXIV:1403.3108;%%
  %9 citations counted in INSPIRE as of 28 gen 2015

\bibitem{ATLASDijet} ATLAS Collaboration, 
  ``Search for New Phenomena in the Dijet Mass Distribution updated
  using 13$\,\text{fb}^{-1}$ of pp Collisions at $\sqrt{s} = 8 \TeV$ collected
  by the ATLAS Detector,''
  ATLAS-CONF-2012-148.

\bibitem{CMSDijet} CMS Collaboration, 
  ``Search for Narrow Resonances using the Dijet Mass Spectrum with
  $19.6\,\text{fb}^{-1}$ of pp Collisions at $\sqrt{s}=8 \TeV$,'' 
  CMS-PAS-EXO-12-059.

\bibitem{AlcarazMaestre:2012vp}
 J.~Alcaraz Maestre {\it et al.}  [SM AND NLO MULTILEG and SM MC Working Groups Collaboration],
 %``The SM and NLO Multileg and SM MC Working Groups: Summary Report,''
 arXiv:1203.6803 [hep-ph].
 %%CITATION = ARXIV:1203.6803;%%

\bibitem{public-code}
  M.~Cacciari, G.~P.~Salam and G.~Soyez,
  public code for validation of a subset of the results in this paper,
  \url{https://github.com/npctests/1404.7353-validation}.
\end{thebibliography}
\end{document}